\documentclass[11pt]{article}

\usepackage[margin=1in]{geometry}
\usepackage[T1]{fontenc}
\usepackage[utf8]{inputenc}
\usepackage{lmodern}
\usepackage{amsmath,amssymb,amsfonts,amsthm}
\usepackage{bm}
\usepackage{mathtools}
\usepackage{physics}
\usepackage{bbm}
\usepackage{graphicx}
\usepackage{tikz}
\usetikzlibrary{arrows.meta,calc,positioning}
\usepackage{caption}
\usepackage{subcaption}
\usepackage{color}
\usepackage{hyperref}
\usepackage[numbers,sort&compress]{natbib}

\hypersetup{
    colorlinks=true,
    linkcolor=blue,
    citecolor=blue,
    urlcolor=blue
}

\newcommand{\ii}{\mathrm{i}}
\newcommand{\Zpart}{\mathcal{Z}}
\newcommand{\vphi}{\varphi}

\newcommand{\Sact}{S}

\newcommand{\D}{\mathcal{D}}

\title{A Convergent Kinetic-Term Perturbation Expansion for $\phi^4$ Theory}

\author{
  Eugene Chen\thanks{Email: yuc090@ucsd.edu} \\
  {\small Department of Physics, University of California San Diego, La Jolla, USA}
}

\date{} 

\begin{document}

\maketitle

\begin{abstract}
We revisit scalar $\phi^4$ theory and construct a reorganized perturbative expansion in which
the kinetic operator, rather than the quartic interaction, is treated as the perturbation.
Starting from the exactly solvable $0$-dimensional model, we show that the resulting series
is convergent for positive coupling and can be written as an expansion in negative powers of
the quartic coupling $\lambda$. We extend the construction to higher-dimensional field theory using
an auxiliary field, and we formulate a discrete lattice version in which multi-site contributions
are systematically organized. We explicitly compute the leading terms in the expansion, study
their continuum limit, and compare against brute-force numerical evaluations of the partition
function. We discuss the relation of this expansion to standard weak-coupling perturbation
theory, strong-coupling expansions, and resummation techniques, and we outline possible
applications to nonperturbative studies of scalar field theories.
\end{abstract}

\tableofcontents
\section{Introduction}

Perturbative approaches to interacting scalar field theories are traditionally organized as
weak-coupling expansions in the interaction strength.  In $\phi^4$ theory, this leads to a
power series in the quartic coupling multiplying Feynman diagrams of increasing loop order.
It has long been understood, however, that such weak-coupling expansions are generically
asymptotic rather than convergent.  Dyson’s argument implies that the radius of convergence of
perturbation theory in the coupling must vanish for interacting theories, and explicit
analyses have shown that perturbative coefficients in scalar $\phi^4$ theory grow
factorially at large order.  Although the weak-coupling expansion for positive coupling is
Borel summable, its asymptotic character \cite{dyson1952divergence,lipatov1977divergence} limits its practical applicability at intermediate
and strong coupling and motivates the search for alternative perturbative frameworks
\cite{bender1969anharmonic,BenderCooperGuralnikSharp1979,benzi1978high}.

In this work, we explore a qualitatively different perturbative organization of $\phi^4$
theory.  Rather than expanding in the interaction strength or reorganizing the weak-coupling
series, we treat the quartic interaction exactly at the level of the path integral and
introduce the kinetic operator as a perturbation.  Importantly, the resulting expansion is
not an expansion about the infinite-coupling limit, nor does it rely on resummation or
optimization of an underlying asymptotic series.  Instead, convergence arises as a
structural consequence of the reorganization itself.

In zero dimensions, this construction coincides algebraically with the standard
strong-coupling expansion familiar from lattice field theory, but admits a particularly
transparent and fully explicit representation.  In this setting, the expansion yields a
genuinely convergent series for all positive coupling, with coefficients that are
sufficiently suppressed at large order to ensure convergence.  The zero-dimensional model
therefore provides a controlled testing ground in which the mechanism underlying convergence
can be analyzed explicitly.

We then extend this construction to higher-dimensional $\phi^4$ theory by introducing an
auxiliary field that renders the integration over the original scalar field local.  After
integrating out the field $\phi(x)$ exactly at each point, the kinetic operator appears as a
set of nonlocal multi-site interactions for the auxiliary field, and the perturbative
expansion becomes an expansion in powers of the kinetic term.  On a discrete lattice, these
contributions naturally organize into one-site, two-site, and higher multi-site terms, whose
structure we compute explicitly.  For small lattice sizes, we compare truncated versions of
the resulting expansion to brute-force numerical evaluations of the lattice path integral
and find good agreement.

The expansion developed here differs conceptually from both traditional strong-coupling
series and variational or optimized perturbative schemes.  It involves no variational
parameters or optimization criteria and does not rely on the resummation of an asymptotic
series.  Instead, the interaction term is treated exactly from the outset, and the
perturbative expansion is organized solely in powers of the kinetic operator.

For any finite lattice system, this construction yields a term-by-term convergent expansion
in arbitrary spacetime dimension.  Moreover, convergence persists as the lattice size is
taken arbitrarily large.  The zero-dimensional theory provides a particularly transparent
setting in which convergence can be demonstrated analytically, but the same mechanism
extends to higher-dimensional lattice field theory at finite and arbitrarily large volume.

Any potential subtleties in the continuum limit are therefore not associated with the
divergence of the perturbative series itself, but rather with the definition of the
continuum theory and the behavior of individual terms in that limit.  In this sense, the
present construction shows that asymptoticity is not an intrinsic feature of perturbative
expansions in interacting field theories, but a consequence of the chosen organization of
the expansion.

A notable and somewhat unexpected outcome of this construction is the interplay between
ultraviolet behavior and the convergence of the perturbative expansion. In conventional
perturbation theory, ultraviolet divergences arising from momentum integrals, such as
\[
\int \dd^d k\,\frac{1}{k^2+m^2} \sim \Lambda^{d-2},
\]
are usually treated as independent from the large-order divergence of the perturbative
series. In the present framework, this separation no longer holds. Requiring a genuinely
convergent expansion forces explicit factors of the lattice spacing $\Delta x$ to appear
in each contribution as a consequence of exact local integration and dimensional analysis.
Since the ultraviolet cutoff scales as $\Lambda \sim 1/\Delta x$, these factors of
$\Delta x$ systematically balance the powers of $\Lambda$ generated by momentum integrals.
Ultraviolet divergences are therefore controlled not by the introduction of counterterms,
but by the same structural mechanism that enforces convergence of the expansion itself.

The remainder of this paper is organized as follows.  In
Section~\ref{sec:standard-perturbation} we briefly review the standard weak-coupling expansion
and its asymptotic behavior.  Section~\ref{sec:0d-kinetic} develops the kinetic-term expansion
in the zero-dimensional theory and analyzes its convergence properties.  In
Section~\ref{sec:continuum-aux} we present the auxiliary-field formulation and the resulting
reorganized expansion in the continuum theory.  Section~\ref{sec:lattice} applies the method
to a discrete lattice and computes one-, two-, and three-site contributions explicitly,
followed by numerical comparisons.  Higher-order diagrams are discussed in
Section~\ref{sec:higher-order}.  Connections to related work and a discussion
of triviality in $\phi^4$ theory are presented in Section~\ref{sec:discussion}.
We conclude in Section~\ref{sec:conclusions} with a discussion
of the implications of this expansion for perturbation theory and possible applications.


\section{Standard perturbation theory for $\phi^4$ theory}
\label{sec:standard-perturbation}

In this section we briefly review the usual weak-coupling perturbation theory for $\phi^4$
theory, highlight its asymptotic nature, and set up notation for the rest of the paper.

\subsection{The $0$-dimensional $\phi^4$ integral and asymptotic series}
\label{subsec:0d-standard}

Consider the standard $0$-dimensional $\phi^4$ partition function,
\begin{equation}
    \Zpart(J)
    = \int_{-\infty}^{\infty} \dd \vphi\;
      \exp\!\left[-\frac{m^2}{2}\vphi^2 - \lambda \vphi^4 + J \vphi\right].
\end{equation}
In the conventional weak-coupling approach, one expands in powers of $\lambda$ around the
Gaussian integral. This leads formally to a double series of the form
\begin{equation}
    \Zpart(J)
    = \sum_{n,n'=0}^{\infty} c_{n,n'}\, \lambda^n \, J^{n'} ,
\end{equation}
whose coefficients exhibit factorial growth at large order, rendering the series
asymptotic rather than convergent.

Explicitly, expanding the interaction and source terms yields
\begin{equation}
    \begin{aligned}
         \Zpart(J)
         &= \int_{-\infty}^{\infty} \dd\vphi \, e^{-\frac{m^2}{2}\vphi^2}
            \sum_{n,n'=0}^\infty
            \frac{(-\lambda\vphi^4)^n (J\vphi)^{n'}}{n!\,n'!} \\
         &= \sum_{n,n'=0}^\infty
            \frac{(-1)^n}{n!\,(2n')!}
            \Gamma\!\left(2n+n'+\frac{1}{2}\right)
            \left(\frac{2}{m^2}\right)^{2n+n'+\frac{1}{2}}
            \lambda^n J^{2n'} ,
    \end{aligned}
\end{equation}
where only even powers of $J$ survive due to the $\vphi \to -\vphi$ symmetry of the
integrand. The expansion coefficients are therefore
\begin{equation}
    c_{n,n'}
    = \frac{(-1)^n}{n!\,(2n')!}
      \Gamma\!\left(2n+n'+\frac{1}{2}\right)
      \left(\frac{2}{m^2}\right)^{2n+n'+\frac{1}{2}} .
\end{equation}
For fixed $n'$ and large $n$, these coefficients scale as
\begin{equation}
    c_{n,n'} \sim \frac{(2n)!}{n!},
\end{equation}
demonstrating the factorial growth characteristic of an asymptotic perturbative
series.

\subsection{Higher-dimensional $\phi^4$ theory and Feynman diagrams}
\label{subsec:d-standard}

In $d$ spacetime dimensions, the Euclidean action for a real scalar field with quartic
self-interaction is
\begin{equation}
    \Sact[\phi]
    = \int \dd^d x \left[
      \frac{1}{2} (\partial_\mu \phi)(\partial^\mu \phi)
      + \frac{m^2}{2} \phi^2
      + \lambda \phi^4
      \right].
\end{equation}
Correlation functions and the partition function are conventionally computed by expanding
around the free Gaussian theory in powers of the coupling $\lambda$.

The resulting perturbative expansion is organized in terms of Feynman diagrams constructed
from the free propagator and quartic interaction vertices. While this diagrammatic expansion
is systematically improvable order by order in $\lambda$, the associated series is
asymptotic rather than convergent, with coefficients that grow factorially at large order.
As a consequence, the weak-coupling expansion does not provide direct quantitative control
at finite or moderately strong coupling.

The interpretation of such divergent perturbative series therefore typically relies on
additional resummation techniques, such as Borel summation or resurgent analysis, which aim
to reconstruct finite results from the asymptotic expansion.
In favorable cases, when the theory is Borel summable, these procedures can uniquely 
reproduce the exact result from perturbation theory. However, they generally require 
detailed information about high-order coefficients and do not by themselves yield a 
rapidly convergent or practically useful expansion at moderate coupling.

In contrast, the approach explored in this work reorganizes the path integral itself rather
than post-processing a divergent perturbative series. By treating the quartic interaction
nonperturbatively and introducing the kinetic operator as a perturbation, this framework
leads to alternative expansions with improved convergence properties, providing a different
starting point for systematic calculations beyond the conventional weak-coupling expansion.

\section{Kinetic-term perturbation in $0$ dimensions}
\label{sec:0d-kinetic}

We now turn to the central construction of this work: a perturbative expansion in which
the quartic interaction is treated exactly, while the mass term plays the role of a
perturbation. In $0$ dimensions, this can be implemented in a particularly transparent way.

\subsection{Exact reorganization of the $0$-dimensional partition function}
\label{subsec:0d-reorg}

We start again from the generating functional
\begin{equation}
    \Zpart(J)
    = \int_{-\infty}^{\infty} \dd \vphi\;
      \exp\!\left[-\frac{m^2}{2}\vphi^2 - \lambda \vphi^4 + J \vphi\right],
\end{equation}
and reorganize the perturbative expansion by treating the Gaussian term as a
perturbation around the pure quartic theory.
Expanding the exponential in powers of $m^2$ and $J$, we obtain
\begin{equation}
    \Zpart(J)
    = \int_{-\infty}^{\infty} \dd \vphi\;
      e^{-\lambda \vphi^4}
      \sum_{n,n'=0}^{\infty}
      \frac{(-\tfrac{m^2}{2})^{n} J^{n'}\, \vphi^{2n+n'}}{n!\,n'!}.
\end{equation}

Since the weight $e^{-\lambda\vphi^4}$ is an even function of $\vphi$, only even
powers of $n'$ contribute to the integral.
Writing $n' \to 2n'$, the $\vphi$–integrals can be evaluated exactly using
\begin{equation}
    \int_{0}^{\infty} \dd \vphi\;
    \vphi^{p} e^{-\lambda \vphi^{4}}
    = \frac{1}{4}\,\lambda^{-\frac{p+1}{4}}
      \Gamma\!\left(\frac{p+1}{4}\right),
\end{equation}
which yields the exact double series
\begin{equation}
\label{eq:Z0d-strong}
    \Zpart(J)
    = \sum_{n,n'=0}^{\infty}
      \frac{(-\tfrac{m^2}{2})^{n} J^{2n'}}{2\,n!\,(2n')!}\,
      \lambda^{-\alpha(n,n')}
      \Gamma\!\left(\alpha(n,n')\right),
\end{equation}
with
\begin{equation}
    \alpha(n,n') = \frac{n+n'}{2} + \frac{1}{4}.
\end{equation}

Equation~\eqref{eq:Z0d-strong} provides an exact reorganization of the partition
function as a strong-coupling expansion in inverse powers of $\lambda$.
Unlike the standard perturbative series in $\lambda$, this expansion has
coefficients controlled by Gamma functions and is convergent for all
$\lambda > 0$, illustrating explicitly how treating the interaction exactly
alters the analytic structure of the expansion.

Parts of the techniques used in this work may be familiar from the strong-coupling expansion
literature. However, we avoid using the term "strong-coupling expansion" for two reasons.
First, strong-coupling expansions comprise a broader class of techniques, many of which are
orthogonal to the present construction. Second, the terminology can be misleading in the
present context. Because the path integral samples arbitrarily large field amplitudes, the
quartic term $\lambda \phi^4$ controls the large-field behavior of the integrand for any
$\lambda>0$. As a result, treating $\lambda \phi^4$ perturbatively is intrinsically
non-uniform over field space and leads to factorial growth of perturbative coefficients.
In contrast, expanding the lower-order quadratic term preserves uniform control of the
large-field region and yields a convergent expansion for all $\lambda>0$.

\subsection{Numerical check in $0$ dimensions}
\label{subsec:0d-numerics}

To verify explicitly that the reorganized expansion derived in
Sec.~\ref{subsec:0d-reorg} converges to the exact result, we evaluate the
$0$-dimensional generating functional numerically and compare it to partial sums
of Eq.~\eqref{eq:Z0d-strong}.

The exact partition function $Z(J)$ is computed by direct numerical quadrature of
the defining integral in Eq.~\eqref{eq:Z0d-strong} over the real line.
We use adaptive Gauss--Kronrod integration on the domain $(-\infty,\infty)$, which
is numerically stable due to the rapid exponential decay of the integrand ensured
by the quartic interaction.
Throughout this section we fix $m^2=1$ and $\lambda=1$, and evaluate $Z(J)$ for a
range of source values $J$.

The reorganized expansion is evaluated by truncating the double sum in
Eq.~\eqref{eq:Z0d-strong} at finite values $(n_{\max},n'_{\max})$.
This yields a sequence of partial sums that can be directly compared to the exact
quadrature result.

\begin{figure}[t]
    \centering
    \includegraphics[width=0.6\textwidth]{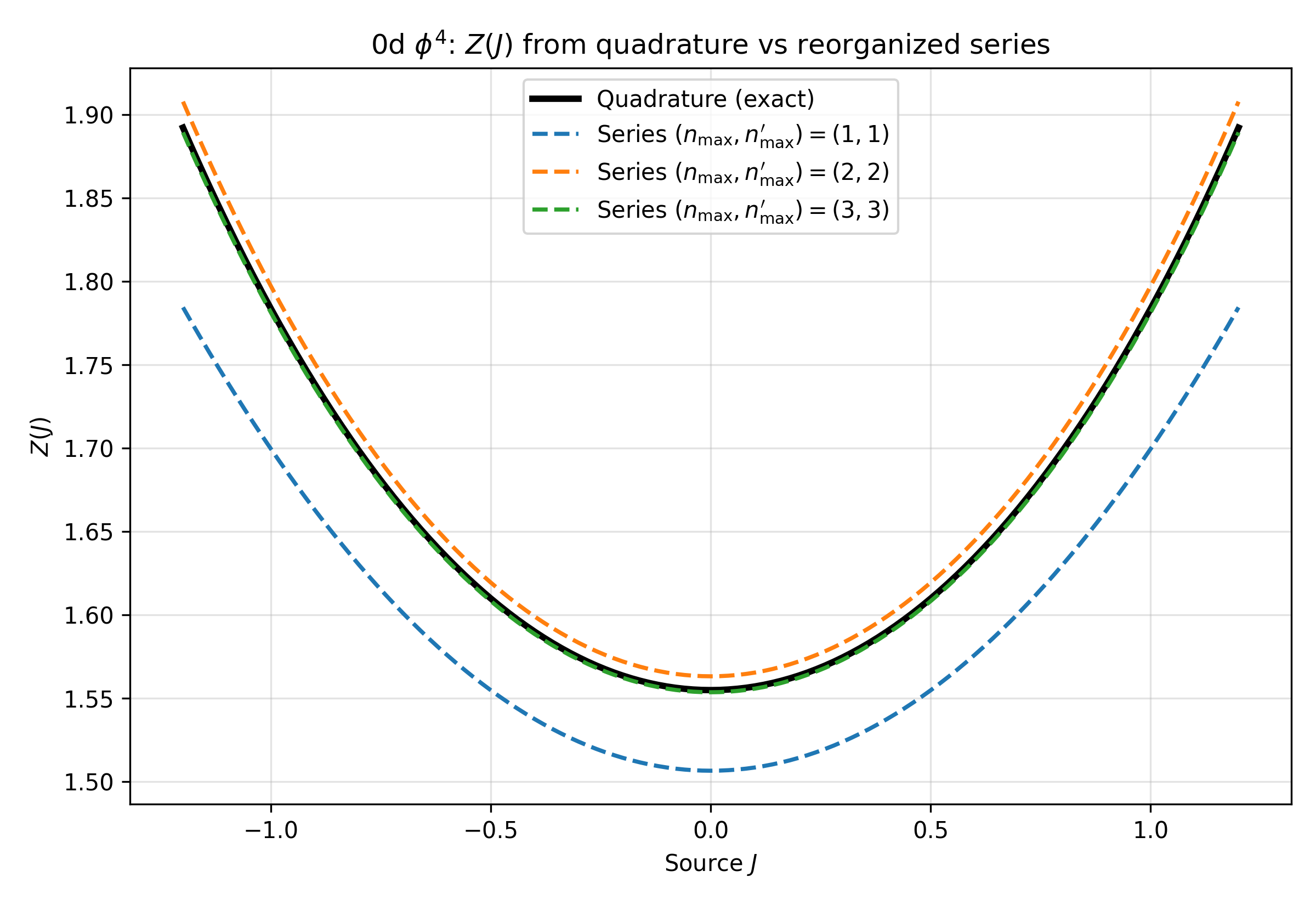}
    \caption{
     Partition function $Z(J)$ of the $0$-dimensional $\phi^4$ theory as a function of the
source $J$.
The solid black curve shows the exact result obtained by direct numerical quadrature,
while the dashed colored curves show partial sums of the reorganized expansion for
increasing truncation orders $(n_{\max},n'_{\max})$.
The uniform convergence of the truncated series to the exact result provides a direct
numerical confirmation of the convergent nature of the reorganized expansion.
    }
    \label{fig:ZJ-vs-J}
\end{figure}

Figure~\ref{fig:ZJ-vs-J} shows the partition function $Z(J)$ as a function of the
source $J$.
The solid curve corresponds to the exact numerical result obtained by quadrature,
while the colored curves show truncated series approximations for increasing
truncation orders $(n_{\max},n'_{\max})$.
As the truncation order is increased, the series curves converge uniformly to the
exact result over the entire plotted range of $J$.
This demonstrates that the reorganized expansion converges to the full generating
functional, rather than merely reproducing a finite set of moments.

Correlation functions may be obtained either by numerical differentiation of
$Z(J)$ or, equivalently, from the coefficients of the corresponding powers of $J$
in Eq.~\eqref{eq:Z0d-strong}.
In particular, the two-point function $\langle\phi^2\rangle$ is extracted directly
from the coefficient of the $J^2$ term.
In all cases we find excellent agreement with direct quadrature, with the same
rapid convergence as a function of truncation order observed for the partition
function.

\section{Continuum $\phi^4$ theory with an auxiliary field}
\label{sec:continuum-aux}

We now generalize the construction to higher-dimensional field theory by introducing
an auxiliary field, which allows us to integrate out the original scalar field locally.
The resulting formulation leads to a convergent reorganization of the path integral
that is valid for all positive coupling and whose exact content is most transparently
understood at the level of resummed series rather than through a closed-form effective
action.

\subsection{Introducing the auxiliary field}
\label{subsec:aux-intro}

We begin with the generating functional of continuum $\phi^4$ theory in the presence of an
external source $J$,
\begin{equation}
    \Zpart[J]
    = \int \D \phi\;
      \exp\!\left[
      -\frac{1}{2} \int \dd^d x \, \phi(x) (-\partial^2 + m^2) \phi(x)
      - \lambda \int \dd^d x\, \phi(x)^4
      + \int \dd^d x\, J(x) \phi(x)
      \right].
\end{equation}
To reorganize the path integral, we introduce an auxiliary (Hubbard--Stratonovich) field
$X(x)$ that linearizes the quadratic term in $\phi$,
\begin{equation}
    \exp\!\left[-\frac{1}{2}\int \dd^d x\, \phi(-\partial^2 + m^2)\phi\right]
    =
    \int \D X\;
    \exp\!\left[
    -\frac{1}{2}\int \dd^d x\, X(-\partial^2 + m^2)^{-1} X
    + \ii \int \dd^d x\, X \phi
    \right].
\end{equation}
With this transformation, the field $\phi(x)$ appears only algebraically and without
derivatives, so the functional integral over $\phi$ becomes ultra-local in position space.

Writing $A \equiv (-\partial^2 + m^2)$, the generating functional may be expressed as
\begin{equation}
\Zpart[J]
= \int \D X\;
\exp\!\left[
-\frac12 \int \dd^d x\,\dd^d y\;
X(x)\,A^{-1}(x-y)\,X(y)
\right]
\prod_x \mathcal I\!\big(\ii X(x)+J(x)\big),
\end{equation}
where the local factor at each spacetime point $x$ is defined by
\begin{equation}
\mathcal I\!\big(X(x),J(x)\big)
\equiv
\int_{-\infty}^{\infty}\dd\phi_x\;
\exp\!\left[
\Delta x^d\Big(-\lambda\,\phi_x^4 + [\ii X(x)+J(x)]\,\phi_x\Big)
\right].
\end{equation}
This integral is closely analogous to the zero-dimensional $\phi^4$ path integral discussed
in Sec.~\ref{subsec:0d-reorg}, with the crucial difference that no quadratic term in $\phi_x$ is
present.

The integral $\mathcal I$ can be evaluated exactly by expanding the linear term and using
parity. Defining $\eta_x \equiv \ii X(x)+J(x)$, one finds
\begin{equation}
\mathcal I(\eta_x)
=
\frac{1}{2}
\sum_{n=0}^{\infty}
\frac{\Gamma\!\left(\frac{2n+1}{4}\right)}{(2n)!}\,
(\Delta x^d)^{\frac{3(2n)+1}{4}}\,
\lambda^{-\frac{2n+1}{4}}\,
\eta_x^{2n}.
\end{equation}
Factoring out the value at $\eta_x=0$,
\begin{equation}
\mathcal I(0)
=
\frac{1}{2}
(\Delta x^d\lambda)^{-1/4}
\Gamma\!\left(\frac14\right),
\end{equation}
we introduce the normalized local functional
\begin{equation}
\widehat{\mathcal I}(\eta_x)
\equiv
\frac{\mathcal I(\eta_x)}{\mathcal I(0)}
=
1+
\sum_{n=1}^{\infty}
\frac{1}{(2n)!}
\frac{\Gamma\!\left(\frac{n}{2}+\frac14\right)}{\Gamma\!\left(\frac14\right)}
\big[(\Delta x^d)^{3/4}\lambda^{-1/4}\eta_x\big]^{2n}.
\end{equation}

\begin{figure}[t]
    \centering
    \includegraphics[width=0.75\textwidth]{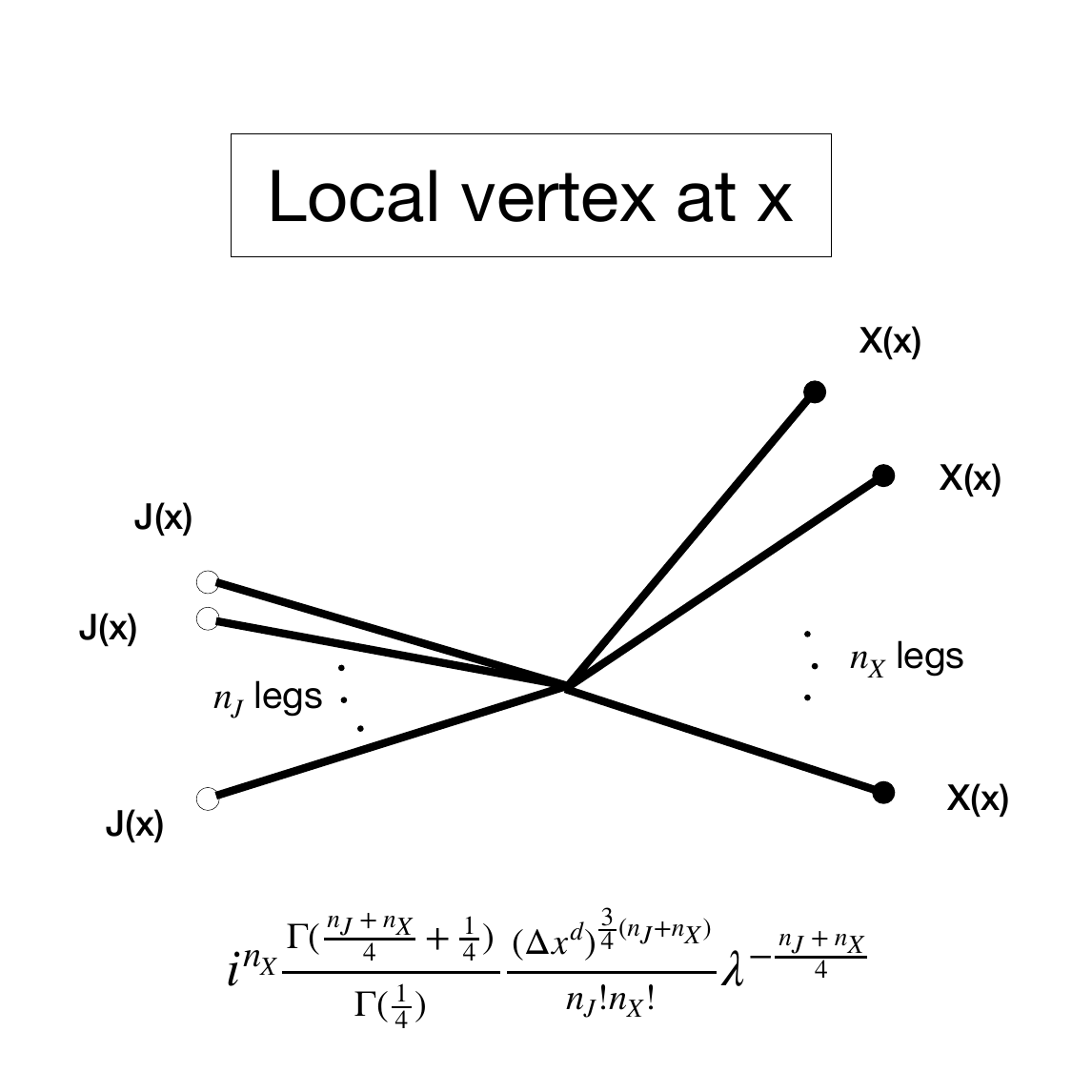}
    \caption{
    Diagrammatic representation of the local factor
    $\widehat{\mathcal I}(\ii X(x)+J(x))$ obtained after integrating out
    the field $\phi(x)$ exactly.
    The vertex corresponds to a single spacetime point $x$ and carries
    arbitrary numbers of insertions of the external source $J(x)$ (open circles)
    and the auxiliary field $X(x)$ (filled circles).
    The associated weight is determined by the exact local integral
    $\widehat{\mathcal I}$ and is symmetric under permutations of the legs.
    }
    \label{fig:local-vertex}
\end{figure}

The generating functional may therefore be written in the compact form
\begin{equation}
\Zpart[J]
=
\frac{1}{\mathcal{N}}
\int \D X\;
\exp\!\left[
-\frac12 X A^{-1} X
\right]
\prod_x \widehat{\mathcal I}\!\big(\ii X(x)+J(x)\big),
\end{equation}
with $\mathcal{N}^{-1} \equiv \prod_x \mathcal I(0)$.
The local structure of the factor $\widehat{\mathcal I}(\ii X(x)+J(x))$ may be represented
diagrammatically as a single vertex with arbitrary insertions of the auxiliary field $X(x)$
and the source $J(x)$, as shown in Fig.~\ref{fig:local-vertex}.

More explicitly, a local vertex with $n_J$ insertions of the source $J(x)$ and
$n_X$ insertions of the auxiliary field $X(x)$ carries a coefficient proportional to 
\begin{equation}
        \frac{1}{n_J! n_X!}\,
        \frac{\Gamma\!\left(\frac{n_J+n_X}{4}+\frac{1}{4}\right)}{\Gamma\!\left(\frac14\right)},
        \qquad (n_J+n_X\ \text{even}),
\end{equation}
as inherited directly from the ultra-local integral
$\widehat{\mathcal I}(\ii X(x)+J(x))$.
When $2n$ auxiliary-field legs are contracted pairwise using the Gaussian measure,
the $n$ Wick contractions generate factorial combinatorial factors.
These are more than compensated by the factorial suppression from the vertex normalization,
so that the net contribution of diagrams involving vertices with large $n_X$ is factorially suppressed.

\subsection{Local expansion and diagrammatic organization}
\label{subsec:aux-expansion}

Rather than rewriting the product over spacetime points in terms of an effective action
involving $\ln \widehat{\mathcal I}$, we work directly with the product representation.
This choice preserves the explicit moment structure of the local integral and makes the
diagrammatic organization of the expansion transparent.

Expanding the normalized local functional,
\begin{equation}
\widehat{\mathcal I}(\eta_x)
=
1
+
\sum_{n=1}^{\infty}
c_n\,
\big[(\Delta x^d)^{3/4}\lambda^{-1/4}\eta_x\big]^{2n},
\qquad
c_n
=
\frac{1}{(2n)!}
\frac{\Gamma\!\left(\frac{n}{2}+\frac14\right)}{\Gamma\!\left(\frac14\right)},
\end{equation}
the product over sites may be written as
\begin{equation}
\prod_x \widehat{\mathcal I}\!\big(\ii X(x)+J(x)\big)
=
1
+
\int \dd^d x\, \mathcal V_x
+
\frac{1}{2!}
\int \dd^d x\,\dd^d y\, \mathcal V_x \mathcal V_y
+\cdots
\label{eq: site expansion}
\end{equation}

where each $\mathcal V_x$ denotes a local polynomial in $X(x)$ and $J(x)$ with explicitly
known coefficients determined by the expansion above.

\begin{figure}[t]
    \centering
    \includegraphics[width=0.55\textwidth]{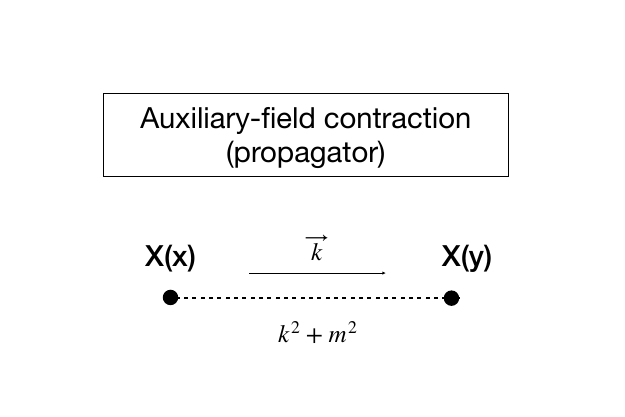}
    \caption{
    Contraction (Gaussian propagator) of the auxiliary field $X$.
    Since $X$ appears quadratically, its two-point function is given by the
    inverse kinetic operator
    $A^{-1}(x-y) = (-\partial^2 + m^2)^{-1}(x-y)$.
    Dashed lines denote contractions of auxiliary-field insertions generated
    by the Gaussian measure.
    }
    \label{fig:aux-propagator}
\end{figure}

Substituting this expansion into the functional integral yields a systematic series
representation of the generating functional in terms of Gaussian correlators of the
auxiliary field $X$. All nonlocal structure arises solely from contractions of auxiliary
fields with the kernel $A^{-1}(x-y)$.
The corresponding contraction rule is shown diagrammatically in
Fig.~\ref{fig:aux-propagator}.

At fixed order, individual terms in this expansion involve momentum integrals that are
ultraviolet divergent if interpreted term by term. As in ordinary perturbation theory,
the expansion is therefore not meant to be evaluated diagram by diagram. Instead,
physically meaningful quantities are obtained only after resumming specific infinite
subclasses of diagrams. In particular, repeated insertions of the lowest-order local
two-$X$ vertex form geometric series that reconstruct finite propagators. An explicit
example of such a resummation for the two-point function is presented in
Appendix~\ref{app:effective-propagator}.

\subsection{Limits $\lambda \to 0$ and $\lambda \to \infty$}
\label{subsec:continuum-limits}

The behavior of the reorganized expansion in the weak- and strong-coupling limits follows
directly from the structure of the resummed propagator.

In the limit $\lambda\to0$, individual terms in the reorganized expansion are singular and do not
admit a smooth limit when considered in isolation. However, the expansion is uniformly
convergent for all $\lambda>0$, and the resummation of the leading geometric series yields an
effective propagator of the form
\begin{equation}
\frac{1}{k^2+m^2+\delta m^2(\lambda)} 
\label{effective propagator}
\end{equation}
with $\delta m^2(\lambda)\propto \lambda^{1/2}$. A technical subtlety arises here: while the power series representation of this propagator has a finite radius of convergence in k-space, $|k^2+m^2|<\delta m^2(\lambda)$, the resummed expression eq.~\ref{effective propagator} provides its unique analytic continuation to the entire momentum domain. As detailed in Appendix.~\ref{app:effective-propagator}, this continuation preserves the analyticity of the propagator across all scales and ensures the smooth recovery of the free-theory results as $\lambda\to 0$.

In the opposite limit $\lambda\to\infty$, all local vertices generated by $\widehat{\mathcal I}$
are parametrically suppressed, and the auxiliary-field theory becomes purely Gaussian.
The effective mass grows with $\lambda$, leading to short-range correlations and a decoupling
behavior consistent with strong coupling.

Together, these limits show that the auxiliary-field formulation provides an exact,
convergent reorganization of continuum $\phi^4$ theory in which both weak- and
strong-coupling regimes are captured within a single unified framework.

\section{Discrete lattice formulation}
\label{sec:lattice}

To test the method in a controlled setting and prepare for possible numerical applications,
we now consider a discrete lattice version of the theory.

\subsection{Lattice action and the matrix $A_{nm}$}
\label{subsec:lattice-A}

We discretize the theory on a one-dimensional periodic lattice with $2N+1$ sites,
lattice spacing $\Delta x$, and coordinates $x_n \equiv n\,\Delta x$ for
$n=-N,\ldots,N$.
The lattice action is
\begin{equation}
    \Sact_{\rm lat}[\{\phi_n\}]
    = \frac{\Delta x}{2} \sum_{n,m=-N}^{N} \phi_n\, A_{nm}\, \phi_m
      + \lambda\, \Delta x \sum_{n=-N}^{N} \phi_n^4 ,
\end{equation}
where the matrix $A_{nm}$ represents the discretized kinetic and mass operator.

The discretization of the derivative operator on a lattice is not unique; one may,
for example, use finite-difference operators or a spectral definition.
Here we adopt a spectral representation, which guarantees translational invariance
and diagonalizability by lattice Fourier modes.

With periodic boundary conditions, the allowed lattice momenta are
\begin{equation}
    k_\ell = \frac{2\pi \ell}{(2N+1)\Delta x},
    \qquad \ell = -N,\ldots,N .
\end{equation}
In this basis, the quadratic operator $-\partial^2 + m_0^2$ is diagonal, with
eigenvalues $k_\ell^2 + m_0^2$.
Transforming back to position space, the matrix $A_{nm}$ is therefore
\begin{equation}
    A_{nm}
    = \frac{1}{2N+1}
      \sum_{\ell=-N}^{N}
      \left( k_\ell^2 + m_0^2 \right)
      e^{ i k_\ell (x_n - x_m) } .
\end{equation}
This form makes explicit that $A_{nm}$ depends only on the site separation
$n-m$, reflecting translational invariance on the lattice.

The inverse matrix $A^{-1}_{nm}$, which appears as the free propagator of the theory,
is given by
\begin{equation}
    (A^{-1})_{nm}
    = \frac{1}{2N+1}
      \sum_{\ell=-N}^{N}
      \frac{ e^{ i k_\ell (x_n - x_m) } }{ k_\ell^2 + m_0^2 } .
\end{equation}
This representation will be particularly useful in later sections, where Gaussian
integrals over the lattice fields are evaluated exactly.

\subsection{One-site contributions to $Z[0]$}
\label{subsec:lattice-1site}

Integrating out the original fields $\phi_n$ locally (at fixed auxiliary field) and
expanding in powers of the kinetic term produces a sum of contributions organized by the
number of \emph{distinct lattice sites} that appear in each term.  The one-site sector is
the part in which all insertions come from a single site $m$, and therefore depends only
on the diagonal element $A_{mm}$.

Starting from the $J=0$ partition function after the local $\phi_n$ integrations, we may
write
\begin{equation}
\begin{aligned}
Z[0]
&=\frac{\Big[\frac{\Gamma(\frac14)}{2}\,(\lambda \Delta x)^{-\frac14}\Big]^{2N+1}}{\mathcal N}
\int \prod_{n=-N}^{N} \dd X_n \;
\exp\!\left[
-\frac12
\sum_{n',m'=-N}^{N}
X_{n'}(A^{-1})_{n'm'}X_{m'}
\right]
\\
&\quad\times
\prod_{m=-N}^{N}
\Bigg[
1+\sum_{a=1}^{\infty} c_a\,(iX_m)^{2a}
\Bigg].
\end{aligned}
\label{eq:Z0-after-phi}
\end{equation}

with coefficients
\begin{equation}
c_a
\equiv
\frac{\Delta x^{\frac{3a}{2}}}{\lambda^{\frac a2}}\;
\frac{1}{(2a)!}\;
\frac{\Gamma\!\left(\frac a2+\frac14\right)}{\Gamma\!\left(\frac14\right)}.
\label{eq:ca-def}
\end{equation}
Expanding the product in \eqref{eq:Z0-after-phi}, the one-site contribution is obtained by
selecting a single factor $(iX_m)^{2a}$ from one site and $1$ from all other sites, i.e.
\begin{equation}
    \begin{aligned}
        &Z[0]\\
&=\Big[\tfrac{\Gamma(\frac14)}{2}(\lambda\Delta x)^{-\frac14}\Big]^{2N+1}
\Bigg[
1+\sum_{m=-N}^{N}\sum_{a=1}^{\infty} c_a\;
\frac{1}{\mathcal N}\int \prod_{n=-N}^{N} dX_n\;
e^{-\frac12 X\cdot A^{-1}\cdot X}\,(iX_m)^{2a}
\;+\;\text{(two-site)}+\cdots
\Bigg]
\label{eq:Z0-onesite-schematic}
    \end{aligned}
\end{equation}

The required Gaussian moments follow from Wick’s theorem.  For a centered Gaussian with
covariance $A$, one has
\begin{align}
\int \prod_{n=-N}^{N} dX_n\;
e^{-\frac12 X\cdot A^{-1}\cdot X}\;
X_m^{2a}
&=(2\pi)^{\frac{2N+1}{2}}(\det A)^{\frac12}\;
\frac{(2a)!}{2^a a!}\;
(A_{mm})^{a}
=\mathcal N\;
\frac{(2a)!}{2^a a!}\;(A_{mm})^{a}.
\label{eq:gaussian-moment}
\end{align}
On a translationally invariant periodic lattice, $A_{mm}$ is independent of $m$ and can be
written in momentum space as
\begin{equation}
A_{mm}
=\frac{1}{2N+1}\sum_{\ell=-N}^{N}\big(k_\ell^2+m_0^2\big),
\qquad
k_\ell=\frac{2\pi\ell}{(2N+1)\Delta x}.
\label{eq:Amm-momentum}
\end{equation}
It is convenient to factor out $\Delta x^{-2}$ by defining $\tilde m^2\equiv m_0^2\Delta x^2$.
Using
\begin{equation}
\frac{1}{2N+1}\sum_{\ell=-N}^{N} \ell^2
=\frac{N(N+1)}{3},
\end{equation}
we obtain the explicit diagonal element
\begin{equation}
A_{mm}
=\Delta x^{-2}\left[
\tilde m^2+\frac{4\pi^2}{3}\frac{N(N+1)}{(2N+1)^2}
\right].
\label{eq:Amm-explicit}
\end{equation}

Substituting \eqref{eq:gaussian-moment} into \eqref{eq:Z0-onesite-schematic} and using
$(i)^{2a}=(-1)^a$ and $\sum_m 1 = 2N+1$, the full one-site sector resums to
\begin{equation}
    \begin{aligned}
        &Z[0]\\
&=\Big[\tfrac{\Gamma(\frac14)}{2}(\lambda\Delta x)^{-\frac14}\Big]^{2N+1}
\Bigg[
1+\sum_{a=1}^{\infty} (-1)^a\;
\frac{2N+1}{2^a a!}\;
\frac{\Gamma\!\left(\frac a2+\frac14\right)}{\Gamma\!\left(\frac14\right)}\;
\frac{\Delta x^{\frac{3a}{2}}}{\lambda^{\frac a2}}\;
\big(A_{mm}\big)^{a}
\;+\;\text{(two-site)}+\cdots
\Bigg]
\nonumber\\[4pt]
&=\Big[\tfrac{\Gamma(\frac14)}{2}(\lambda\Delta x)^{-\frac14}\Big]^{2N+1}
\Bigg[
1+\sum_{a=1}^{\infty} (-1)^a\;
\frac{2N+1}{2^a a!}\;
\frac{\Gamma\!\left(\frac a2+\frac14\right)}{\Gamma\!\left(\frac14\right)}\;
\frac{1}{\big(\lambda \Delta x^{3}\big)^{\frac a2}}\;
\left(
\tilde m^2+\frac{4\pi^2}{3}\frac{N(N+1)}{(2N+1)^2}
\right)^{a}\\
&+\;\text{(two-site)}+\cdots
\Bigg].
\label{eq:Z0-onesite-final}
    \end{aligned}
\end{equation}
Equation \eqref{eq:Z0-onesite-final} is the explicit one-site contribution: it depends on
the microscopic parameters only through the dimensionless combination
$\lambda\Delta x^{3}$ and through the diagonal element $A_{mm}$ (equivalently, the momentum
sum in \eqref{eq:Amm-momentum}).

\subsection{Two-site terms}
\label{subsec:lattice-2site}

The next class of contributions arises from selecting nontrivial factors from \emph{two}
distinct lattice sites in the expansion of the product in
Eq.~\eqref{eq:Z0-after-phi}.  These terms probe off-diagonal correlations of the
auxiliary field and therefore depend on $A_{nm}$ with $n\neq m$.

Starting from
\eqref{eq:Z0-after-phi}--\eqref{eq:ca-def}, the two-site sector is obtained by choosing
$(iX_{n})^{2a_1}$ from site $n$ and $(iX_{m})^{2a_2}$ from site $m$, with $n>m$, and $1$
from all other sites.  This yields
\begin{equation}
    \begin{aligned}
        &Z[0]\\
&=\Big[\tfrac{\Gamma(\frac14)}{2}(\lambda\Delta x)^{-\frac14}\Big]^{2N+1}
\Bigg[
1+\text{(one-site)}
+\sum_{n>m}\sum_{a_1,a_2\ge 1} c_{a_1}c_{a_2}\;
\frac{1}{\mathcal N}\int \prod_{\ell=-N}^{N} dX_\ell\;
e^{-\frac12 X\cdot A^{-1}\cdot X}\,
(iX_n)^{2a_1}(iX_m)^{2a_2}\\
&+\text{(three-site)}+\cdots
\Bigg]
\label{eq:Z0-twosite-start}
    \end{aligned}
\end{equation}
where $c_a$ is given in \eqref{eq:ca-def}.  Using $(i)^{2a}=(-1)^a$, we may write the
two-site contribution as
\begin{equation}
\text{Term}_{\rm 2\text{-}site}
=
\sum_{n>m}\sum_{a_1,a_2\ge 1}
(-1)^{a_1+a_2}\,
\frac{\Delta x^{\frac{3(a_1+a_2)}{2}}}{\lambda^{\frac{a_1+a_2}{2}}}\,
\frac{\Gamma(\frac{a_1}{2}+\frac14)\Gamma(\frac{a_2}{2}+\frac14)}{\Gamma(\frac14)^2}\;
I_{nm}^{(a_1,a_2)} ,
\label{eq:Term-2site-master}
\end{equation}
with the normalized Gaussian moments
\begin{equation}
I_{nm}^{(a_1,a_2)}
\equiv
\frac{1}{(2a_1)!(2a_2)!\,\mathcal N}
\int \prod_{\ell=-N}^{N} dX_\ell\;
e^{-\frac12 X\cdot A^{-1}\cdot X}\,
X_n^{2a_1}X_m^{2a_2}.
\label{eq:Inm-def}
\end{equation}

\paragraph{Explicit combinatorial formula.}
The Gaussian integral \eqref{eq:Inm-def} can be evaluated by Wick contractions.  Let
$b$ denote the number of \emph{cross-pairings} between $X_n$ and $X_m$.  Because each
cross-pairing consumes one $X_n$ and one $X_m$, the number of cross-pairings ranges from
$b=0$ to $b=\min(a_1,a_2)$, and contributes a factor $(A_{nm})^{2b}$ (since there are
$2b$ cross-contracted fields in total).  The remaining fields are paired within each site,
giving factors $A_{nn}^{a_1-b}$ and $A_{mm}^{a_2-b}$.  The resulting exact combinatorial
sum is
\begin{equation}
I_{nm}^{(a_1,a_2)}
=
\sum_{b=0}^{\min(a_1,a_2)}
\frac{
A_{nn}^{\,a_1-b}\,
A_{mm}^{\,a_2-b}\,
A_{nm}^{\,2b}
}{
2^{\,a_1+a_2-2b}\,
(2b)!\,
(a_1-b)!\,
(a_2-b)!
}\,.
\label{eq:Inm-combinatorial}
\end{equation}
This formula makes manifest that the two-site sector depends only on the covariance
elements $A_{nn}$, $A_{mm}$, and $A_{nm}$.

\paragraph{Translation-invariant form.}
For the periodic lattice considered here, $A_{nm}$ depends only on the separation
$r\equiv n-m$, so it is convenient to define
\begin{equation}
A_{nm} \equiv A(r),\qquad A_{nn}=A(0),
\end{equation}
and rewrite the site sum as a sum over separations:
\begin{equation}
\sum_{n>m} f(n-m)
=\sum_{r=1}^{2N} (2N+1-r)\, f(r).
\label{eq:pair-sum-separation}
\end{equation}
Using \eqref{eq:Inm-combinatorial} and $A_{nn}=A_{mm}=A(0)$, we obtain
\begin{align}
\sum_{n>m} I_{nm}^{(a_1,a_2)}
&=
\sum_{r=1}^{2N} (2N+1-r)
\sum_{b=0}^{\min(a_1,a_2)}
\frac{
A(0)^{\,a_1+a_2-2b}\,
A(r)^{\,2b}
}{
2^{\,a_1+a_2-2b}\,
(2b)!\,
(a_1-b)!\,
(a_2-b)!
}\,.
\label{eq:sum-Inm-translation}
\end{align}

\paragraph{Momentum-space expressions for $A(0)$ and $A(r)$.}
With the spectral definition of the lattice operator from
Sec.~\ref{subsec:lattice-A}, one has
\begin{equation}
A(r)=\frac{1}{2N+1}\sum_{\ell=-N}^{N}\big(k_\ell^2+m_0^2\big)\,e^{ik_\ell r\Delta x},
\qquad
k_\ell=\frac{2\pi\ell}{(2N+1)\Delta x}.
\label{eq:A-of-r-momentum}
\end{equation}
In particular,
\begin{equation}
A(0)=\Delta x^{-2}\left[
\tilde m^2+\frac{4\pi^2}{3}\frac{N(N+1)}{(2N+1)^2}
\right],
\qquad \tilde m^2\equiv m_0^2\Delta x^2,
\label{eq:A0-explicit}
\end{equation}
while for $r\neq 0$ the sum can be evaluated to the closed form
\begin{equation}
A(r)
=\Delta x^{-2}\;
\frac{(-1)^r\,2\pi^2}{(2N+1)^2}\,
\frac{\cos\!\big(\frac{\pi r}{2N+1}\big)}{\sin^2\!\big(\frac{\pi r}{2N+1}\big)}.
\label{eq:Ar-explicit}
\end{equation}
Equations \eqref{eq:Term-2site-master}--\eqref{eq:Ar-explicit} provide a complete and
explicit expression for the two-site contributions in terms of $\lambda$, $\Delta x$,
$N$, and the lattice covariance elements.

\subsection{Three-site terms and the general pattern}
\label{subsec:lattice-3site}

The three-site sector provides the first genuinely nontrivial realization of the general
multi-site structure of the expansion.  It arises from selecting nontrivial factors from
three distinct lattice sites in the product appearing in
Eq.~\eqref{eq:Z0-after-phi}, and depends on all pairwise covariance elements among the
three sites.

Proceeding as in the one- and two-site cases, the three-site contribution to the partition
function takes the form
\begin{align}
\text{Term}_{\rm 3\text{-}site}
&=
\sum_{n_1>n_2>n_3}
\sum_{a_1,a_2,a_3\ge 1}
(-1)^{a_1+a_2+a_3}
\frac{\Delta x^{\frac{3(a_1+a_2+a_3)}{2}}}{\lambda^{\frac{a_1+a_2+a_3}{2}}}
\frac{
\Gamma(\frac{a_1}{2}+\frac14)
\Gamma(\frac{a_2}{2}+\frac14)
\Gamma(\frac{a_3}{2}+\frac14)
}{
\Gamma(\frac14)^3
}
\,
I^{(a_1,a_2,a_3)}_{n_1n_2n_3},
\label{eq:Term-3site-master}
\end{align}
where the normalized Gaussian moments are defined by
\begin{align}
I^{(a_1,a_2,a_3)}_{n_1n_2n_3}
\equiv
\frac{1}{(2a_1)!(2a_2)!(2a_3)!\,\mathcal N}
\int \mathcal D X\;
e^{-\frac12 X^T A^{-1} X}\;
X_{n_1}^{2a_1} X_{n_2}^{2a_2} X_{n_3}^{2a_3}.
\label{eq:I123-def}
\end{align}

\paragraph{General Wick-contraction structure.}
The evaluation of \eqref{eq:I123-def} follows directly from Wick’s theorem.
Each contraction pattern can be characterized by nonnegative integers
$b_{ij}$ ($i\le j$, $i,j\in\{1,2,3\}$), where
$b_{ii}$ counts the number of self-contractions at site $n_i$ and
$b_{ij}$ ($i<j$) counts the number of cross-contractions between sites $n_i$ and $n_j$.
These integers are constrained by
\begin{equation}
2b_{ii} + \sum_{j\neq i} b_{ij} = 2a_i ,
\qquad i=1,2,3,
\label{eq:bij-constraints}
\end{equation}
and the Gaussian moment can be written compactly as
\begin{equation}
I^{(a_1,a_2,a_3)}_{n_1n_2n_3}
=
\sum_{\{b_{ij}\}\in\mathcal D(a_1,a_2,a_3)}
\frac{
A_{n_1n_1}^{\,b_{11}}
A_{n_2n_2}^{\,b_{22}}
A_{n_3n_3}^{\,b_{33}}
A_{n_1n_2}^{\,b_{12}}
A_{n_1n_3}^{\,b_{13}}
A_{n_2n_3}^{\,b_{23}}
}{
2^{\,b_{11}+b_{22}+b_{33}}\;
b_{11}!\,b_{22}!\,b_{33}!\,
b_{12}!\,b_{13}!\,b_{23}!
}.
\label{eq:I123-combinatorial}
\end{equation}
Here $\mathcal D(a_1,a_2,a_3)$ denotes the finite domain of nonnegative integers satisfying
the constraints \eqref{eq:bij-constraints}.  Equation \eqref{eq:I123-combinatorial} makes
explicit that the three-site sector depends on all pairwise covariance elements among the
three sites.

\paragraph{Translation-invariant form.}
On the periodic lattice, translational invariance allows one to rewrite the sum over site
triples in terms of separations.  Writing $r_1=n_1-n_3$ and $r_2=n_2-n_3$ with
$1\le r_2<r_1\le 2N$, one finds
\begin{align}
\sum_{n_1>n_2>n_3} I^{(a_1,a_2,a_3)}_{n_1n_2n_3}
=
\sum_{r_1=2}^{2N}\sum_{r_2=1}^{r_1-1} (2N+1-r_1)\;
\mathcal I^{(a_1,a_2,a_3)}(r_1,r_2),
\label{eq:3site-separation}
\end{align}
where $\mathcal I^{(a_1,a_2,a_3)}(r_1,r_2)$ is obtained from
\eqref{eq:I123-combinatorial} by substituting
$A_{n_in_j}\to A(|n_i-n_j|)$.
In terms of the momentum sums $S_r$ defined in the previous subsection, this expression
can be written explicitly as a finite sum over contraction numbers, with powers of
$S_0$, $S_{r_1}$, $S_{r_2}$, and $S_{r_1-r_2}$.

\paragraph{General pattern.}
The structure of the three-site sector makes the general pattern clear.
At $k$ sites, the contribution involves:
(i) a sum over $k$ independent nonnegative integers $a_i$,
(ii) a finite combinatorial sum over Wick-contraction numbers $b_{ij}$ satisfying linear
constraints,
and (iii) dependence on all pairwise covariance elements $A_{n_in_j}$.
While the number of terms grows rapidly with the number of sites, each sector is finite
and completely determined by Gaussian combinatorics.
This organization provides a systematic and convergent expansion of the lattice partition
function in powers of the kinetic operator.
\subsection{Numerical comparison to brute-force integration}
\label{subsec:lattice-numerics}

For small lattice sizes (e.g.\ $N=1$), it is feasible to evaluate the lattice partition
function
\begin{equation}
    \Zpart_\text{lat}[0]
    = \int \prod_{n=-N}^N \dd \phi_n\;
      \exp\!\left[-\Sact_\text{lat}[\{\phi_n\}]\right]
\end{equation}
directly by numerical integration.
In this subsection, we compare the exact result obtained from brute-force integration
to successive partial sums of the reorganized kinetic-term expansion, including the
one-site, two-site, and three-site contributions derived above.

\begin{figure}[t]
    \centering
    \includegraphics[width=0.6\textwidth]{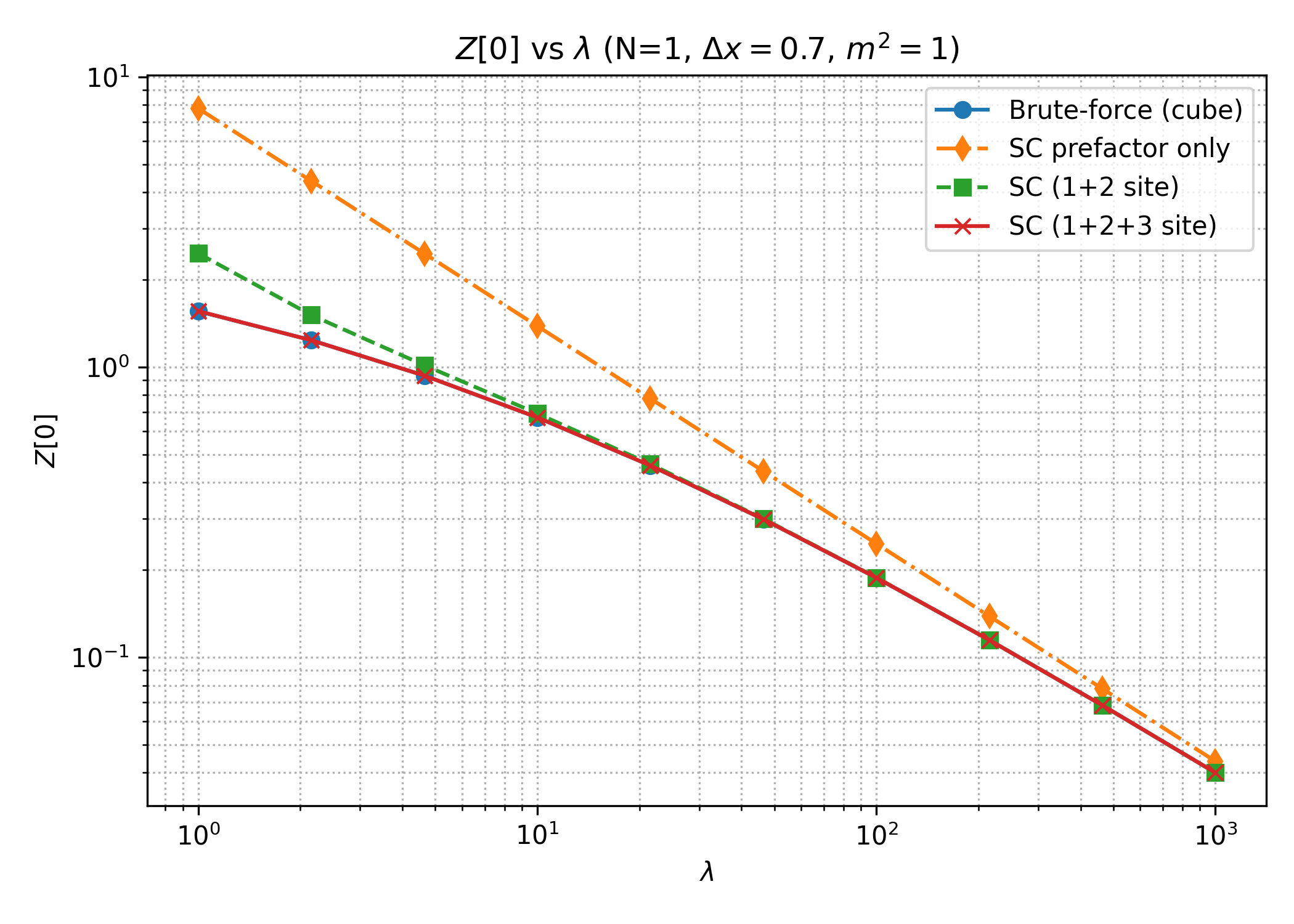}
    \caption{
    Partition function $Z[0]$ of the lattice $\phi^4$ theory as a function of the
    quartic coupling $\lambda$ for $N=1$, $\Delta x=0.7$, and $m^2=1$.
    The blue curve shows the exact result obtained from brute-force numerical integration
    over the lattice fields.
    The remaining curves show successive truncations of the reorganized kinetic-term
    expansion: the strong-coupling prefactor only, the prefactor plus one- and two-site
    contributions, and the prefactor plus one-, two-, and three-site contributions.
    The progressive improvement demonstrates the systematic convergence of the expansion
    as higher multi-site terms are included.
    }
    \label{fig:Z0-lattice}
\end{figure}

Figure~\ref{fig:Z0-lattice} shows the partition function $Z[0]$ as a function of the
quartic coupling $\lambda$ for a lattice with $N=1$, lattice spacing $\Delta x=0.7$,
and mass parameter $m^2=1$.
The brute-force result is obtained by direct numerical integration over the three lattice
fields, while the strong-coupling curves correspond to truncations of the kinetic-term
expansion at increasing site order.

We observe that the leading strong-coupling prefactor alone captures only the qualitative
scaling with $\lambda$.
Including the one-site contributions significantly improves the agreement, while the
addition of two-site terms brings the expansion into near-quantitative agreement with
the exact result over a wide range of couplings.
The inclusion of the three-site sector further refines the approximation and yields
excellent agreement with the brute-force result across the entire range of $\lambda$
shown.
This provides a direct numerical confirmation that the reorganized expansion converges
systematically as higher multi-site contributions are included.

\section{Higher-order diagrams}
\label{sec:higher-order}

Beyond the leading tree-level contributions, the reorganized expansion includes loop
diagrams such as the sunset diagram. We briefly outline their structure and scaling.
\subsection{The sunset diagram}
\label{subsec:sunset}

We now consider the leading nontrivial loop contribution to the two-point function in the
auxiliary-field formulation.  It is convenient to introduce the dimensionless effective coupling
\begin{equation}
    \tilde{\lambda}
    \;\equiv\;
    \lambda\,\Delta x^{-d}
    \left[\frac{2\,\Gamma(\tfrac14)}{\Gamma(\tfrac34)}\right]^2 .
\end{equation}
In terms of this parameter, the resummed auxiliary-field propagator derived in the previous section
may be written as
\begin{equation}
    D_X(k)
    \;=\;
    \tilde{\lambda}^{1/2}
    \left[
    1-\frac{\tilde{\lambda}^{1/2}}{\tilde{\lambda}^{1/2}+m^2+k^2}
    \right].
\end{equation}
For later convenience, we define the dimensionless factor
\begin{equation}
    G(k)
    \;\equiv\;
    1-\frac{\tilde{\lambda}^{1/2}}{\tilde{\lambda}^{1/2}+m^2+k^2}
    \;=\;
    \frac{k^2+m^2}{k^2+m^2+\tilde{\lambda}^{1/2}} .
\end{equation}

\begin{figure}[t]
    \centering
    \includegraphics[width=0.45\textwidth]{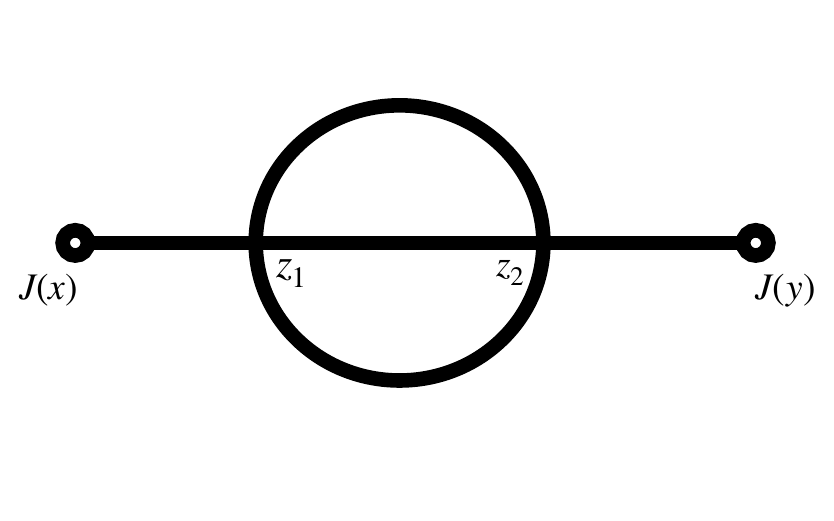}
    \caption{
    Sunset diagram contributing to the two-point function in the auxiliary-field formulation.
    Internal lines represent resummed auxiliary-field propagators, and the two vertices correspond
    to local $JX^3$ insertions generated by the expansion of $\widehat{\mathcal I}$.
    }
    \label{fig:sunset}
\end{figure}

The leading nontrivial diagram contributing to the two-point function is the sunset diagram,
obtained by contracting two local $JX^3$ vertices with three auxiliary-field propagators.
The diagram is shown in Fig.~\ref{fig:sunset}.

Using the vertex coefficient obtained from the local expansion of $\widehat{\mathcal I}$, the
sunset contribution takes the form
\begin{equation}
    \begin{aligned}
        \Pi_{\rm sun}(k)
        \;=\;
        \Delta x^{-2d}
        \left[
            i^3\,
            \frac{\Gamma(\tfrac54)}{\Gamma(\tfrac14)}\,
            \frac{\Delta x^{2d}}{3!\,\tilde{\lambda}}\,
            \left(\frac{2\,\Gamma(\tfrac14)}{\Gamma(\tfrac34)}\right)^2
        \right]^2
        \tilde{\lambda}^{3/2}
        \int\!\frac{d^d k_1}{(2\pi)^d}
             \frac{d^d k_2}{(2\pi)^d}\,
        G(k_1)\,G(k_2)\,G(k-k_1-k_2) .
    \end{aligned}
\end{equation}
The overall factor $\Delta x^{-2d}$ arises from converting the two lattice sums over vertex
positions into continuum integrals, $\sum_x \to \Delta x^{-d}\int d^d x$.

Power counting provides a simple estimate of the ultraviolet behavior of this contribution.
For fixed external momentum $k$, the two-loop momentum integral scales as
\begin{equation}
    \int d^d k_1\, d^d k_2\,
    \frac{1}{k_1^2+\tilde m^2}\,
    \frac{1}{k_2^2+\tilde m^2}\,
    \frac{1}{(k-k_1-k_2)^2+\tilde m^2}
    \;\sim\;
    \Lambda^{2d-6}
    \qquad (d>4),
\end{equation}
where $\Lambda\sim\Delta x^{-1}$ is the ultraviolet cutoff.  Combining this with the explicit
prefactor $\Delta x^{2d}\sim\Lambda^{-2d}$, we find that the sunset contribution is suppressed as
$\Lambda^{-6}$ in the continuum limit for $d>4$.

This suppression is consistent with the irrelevance of interaction-induced loop corrections in
dimensions above four.  A more systematic discussion of dimensional dependence and its relation
to triviality is deferred to Sec.~7.

\subsection{General higher-order structure}
\label{subsec:higher-structure}

The structure of the sunset diagram illustrates the general organization of higher-order
contributions in the kinetic-term expansion.  At arbitrary order, diagrams are constructed from
the same two basic building blocks: local vertices generated by the expansion of
$\widehat{\mathcal I}(iX+J)$, and resummed auxiliary-field propagators obtained by summing
repeated insertions of the lowest-order two-$X$ vertex.

At a given order in the expansion, contributions arise from contracting an arbitrary number of
local vertices with the Gaussian measure for the auxiliary field.  Each diagram is therefore
specified by a choice of vertex multiplicities and a pattern of Wick contractions among the
corresponding $X$ fields.  Topologically, the resulting graphs coincide with the usual Feynman
diagrams of a scalar theory, but with non-polynomial local vertices and propagators that already
encode an infinite resummation of kinetic-term insertions.

On a finite lattice, the expansion is absolutely convergent at every order.  Each local vertex
comes with a coefficient that decays factorially with its valence, while Gaussian moments of the
auxiliary field are finite and explicitly computable.  As a result, all multi-site and multi-loop
contributions are finite, and the full series converges for any positive coupling.

In the continuum limit, individual diagrams generally contain ultraviolet-divergent momentum
integrals if interpreted term by term.  As in the sunset example, meaningful results are obtained
only after resumming the geometric series generated by repeated local insertions, which produces
the effective auxiliary-field propagator.  Once this resummation is performed, higher-loop
diagrams are suppressed by explicit powers of the cutoff in dimensions $d>4$, reflecting the
irrelevance of interaction-induced corrections in that regime.

For dimensions $d\le4$, uniform convergence in momentum space is no longer guaranteed, and the
interplay between resummation and renormalization becomes essential.  The present expansion
nevertheless provides a well-defined and systematically improvable organization of contributions,
with all ultraviolet sensitivity isolated into a small number of resummed structures.

A more detailed discussion of the dimensional dependence of the expansion, its relation to
triviality, and the interpretation of the continuum limit is presented in Sec.~7.

\section{Discussion}
\label{sec:discussion}
\subsection{Related reorganizations of perturbative expansions}
\label{subsec:comparison}

A variety of alternative reorganizations of perturbative expansions in scalar field theory
have been explored across statistical mechanics and quantum field theory, motivated by the
asymptotic character and limited applicability of the conventional weak-coupling expansion.
Below we summarize several representative classes of approaches and clarify how they differ
in their organizing principles and notions of convergence.

\paragraph{Strong-coupling, high-temperature, and hopping expansions.}
One natural alternative is provided by strong-coupling
\cite{BenderCooperGuralnikSharp1979,svaiter2005strong,bernard2019nonlocal}
or high-temperature expansions
\cite{domb2000phase,parisi1975high},
which have been extensively studied in lattice scalar field theory
\cite{benzi1978high,kogut1979introduction}
and statistical mechanics
\cite{kleinert2010converting,li2011effective}.
In these approaches, the interaction term is treated exactly while the kinetic or hopping term
is expanded perturbatively, leading to series organized in powers of the hopping or kinetic
operator, often corresponding—after appropriate rescalings—to inverse powers of the coupling.
Such expansions have been used to extract nonperturbative information and critical behavior,
either analytically\cite{freericks1996strong,campostrini1995strong} or through numerical analyses\cite{butera2001remark} of long high-temperature series and extrapolation procedures.
But they are typically formulated as expansions about the infinite-coupling (or decoupled-site)
limit.  As a result, their convergence properties depend sensitively on the model and observable
under consideration and often require resummation or extrapolation techniques.  Moreover, the
relation between these lattice-based expansions and the continuum limit can be subtle.

\paragraph{Cluster, linked-cluster, and polymer expansions.}
Implementing an expansion in the kinetic or hopping term requires a systematic organization of
the resulting nonlocal contributions.  Cluster, linked-cluster, and polymer expansions provide
a natural framework for this purpose, expressing observables as sums over connected graphs or
polymers whose weights are determined by local interactions and lattice embeddings.\cite{wortis1974linked,luscher1988application,gelfand1996series,reisz1995advanced,procacci1998remark} These
methods ensure extensivity of the free energy and correlation functions and form the technical
backbone of many high-temperature and strong-coupling analyses.

\paragraph{$\delta$-expansion and variational perturbation theory.}
A different class of methods seeks to improve the behavior of perturbative expansions by
reorganizing the weak-coupling series itself.  Examples include optimized perturbation theory\cite{stevenson1981optimized,chiku1998optimized,jakovac2016optimized},
the linear $\delta$-expansion\cite{guida1995convergence,yamada2007delta}, variational perturbation theory\cite{kleinert2006path,sissakian1994variational,sissakian1994ss,solovtsova2013variational}, and order-dependent mappings
\cite{seznec1979summation}.
In these approaches, auxiliary variational parameters are introduced and fixed through
order-by-order optimization, leading to convergent sequences of approximants for certain
observables when the parameters are chosen to scale appropriately with perturbative order.
This mechanism has been shown to reproduce strong-coupling scaling behavior in simple models,
such as the anharmonic oscillator
\cite{JankeKleinert1995,Kleinert1998}.
However, the convergence achieved in this framework relies intrinsically on optimization at
each truncation order and does not correspond to a term-by-term convergent expansion with fixed
coefficients.\cite{weissbach2002high,hamprecht2004variational} Indeed, analyses of the $\delta$-expansion in quantum field theory emphasize that
without the correct strong-coupling scaling prescriptions, the resulting approximants may fail
to converge or converge to an incorrect limit
\cite{HamprechtKleinert2003}.

\paragraph{Resummation and resurgent methods.}
In parallel, divergent perturbative series can be assigned finite values through Borel
summation and related resummation techniques.\cite{le2012large,costin2019resurgent,malbouisson1999non}  These methods provide a mathematically precise interpretation of asymptotic expansions by exploiting information about large-order behavior
and the analytic structure of perturbative coefficients, and have been successfully applied to
scalar field theories and quantum-mechanical models
\cite{SeroneSpadaVilladoro2018,MarinoResurgence,sberveglieri2019renormalization}.
While powerful, such approaches operate a posteriori on an asymptotic weak-coupling
series and do not alter the underlying organization of perturbation theory itself; moreover,
their practical implementation typically requires detailed knowledge of large-order
asymptotics.

\paragraph{Constructive field theory and convergent reorganizations.}
Constructive field theory \cite{baez2014introduction,jaffe2000constructive,rivasseau2000constructive,hancox2017solutions,seong2024central} provides an alternative approach in which quantum field theories are
defined through convergent expansions obtained by reorganizing the functional integral itself,
rather than by resumming an asymptotic perturbative series.  Within this framework, techniques
such as cluster and forest formulas yield absolutely convergent expansions by controlling the
combinatorics of perturbative contributions, as exemplified by the loop vertex expansion
\cite{rivasseau2010loop,rivasseau2018loop,sazonov2025variational}.  While intermediate-field representations are often
used as technical tools, convergence in these constructions arises from a combinatorial
reorganization, in contrast to the present work, where convergence follows from exact local
integration of the interaction.

\paragraph{Auxiliary-field and dual formulations.}
Related strong-coupling and duality-based formulations of scalar lattice field theory have also
been explored, including approaches based on field-space Fourier duality that yield
strong-coupling decompositions of lattice observables
\cite{ignatyuk2022strong}.
Such constructions often focus on vacuum quantities and correlation functions at fixed lattice
spacing.  The present work instead emphasizes an exact local integration and source-based
formulation, which makes the structure of the expansion and its convergence properties
transparent from the outset.

\paragraph{Summary.}
The approaches reviewed above differ in how perturbation theory is reorganized and in the
sense in which convergence is achieved, including expansions about the decoupled-site limit,
order-dependent optimization of auxiliary parameters, and resummation of asymptotic series.
The present work is closely related in spirit to strong-coupling and constructive approaches,
sharing the goal of reorganizing the path integral prior to perturbative expansion, but
differs in its technical implementation and motivation.  In particular, convergence here is
achieved through exact local integration of the interaction, leading to a term-by-term
convergent expansion with fixed coefficients, without invoking optimization procedures or
a posteriori resummation prescriptions.

\subsection{Dimensional dependence and triviality issues}
\label{subsec:dimensional}

The behavior of the reorganized expansion depends sensitively on the spacetime
dimension through simple power-counting considerations.
The effective coupling $\tilde{\lambda}$ introduced in
Sec.~\ref{sec:continuum-aux} carries mass dimension
$[\tilde{\lambda}]=4$, while the source has dimension
$[J(x)] = (d+2)/2$.
Here $\tilde{\lambda}$ should be understood as a lattice-regulated
parameter whose appearance is always accompanied by appropriate powers
of the lattice spacing $\Delta x$.

Dimensional analysis allows us to determine the general scaling of loop
contributions up to dimensionless numerical coefficients.
As an illustrative example, consider terms contributing to the two-point
function at $\mathcal{O}(J^2)$.
Such terms take the schematic form
\begin{equation}
    \int \dd^d x\, \dd^d y\; J(x) J(y)
    \int \frac{\dd^d k}{(2\pi)^d} e^{ik(x-y)}
    \prod_{i=1}^{n_{\rm L}}
    \int \frac{\dd^d k_i}{(2\pi)^d}
    \left[
        \frac{\tilde{\lambda}^{1/2}}{k_i^2 + m^2 + \tilde{\lambda}^{1/2}}
    \right]^{n_{\rm P}}
    (\Delta x^d)^{n_{\rm L}}\, \tilde{\lambda}^{1/2},
\end{equation}
where $n_{\rm L}$ denotes the number of independent loop momentum
integrals and $n_{\rm P}$ the number of internal propagator factors.

Each internal momentum integral is regulated by the lattice spacing,
with an effective ultraviolet cutoff $\Lambda \sim 2\pi/\Delta x$.
Power counting then gives
\begin{equation}
    \int \frac{\dd^d k}{(2\pi)^d}
    \frac{1}{(k^2+m^2)^p}
    \sim
    \begin{cases}
        \Lambda^{d-2p}, & d>2p,\\
        f(m), & d\le 2p,
    \end{cases}
\end{equation}
where $f(m)$ remains finite in the continuum limit.

Let $N$ denote the number of internal vertices.
Consider contractions of
\begin{align}
X_{1}^{2n_1}\ldots X_N^{2n_N}\,
X_{x}^{2n_x-1}\,
X_{y}^{2n_y-1},
\end{align}
with all $n_i\ge2$ (since contributions with $n_i=1$ have already
been resummed into effective propagators).
A necessary condition for the appearance of infrared (IR) divergences is
\begin{align}
d\,n_{\rm L} \le 2\,n_{\rm P}.
\end{align}
Using the combinatorial relations between loops, propagators, and
vertices, this condition can be rewritten as
\begin{align}
d \le \frac{2(2N+3)}{N+2} < 4.
\end{align}

Hence, for spacetime dimensions $d>4$, interaction contributions scale as
\begin{equation}
\Lambda^{n_{\rm L}d-2n_{\rm P}}(\Delta x^d)^{n_{\rm L}}
\;\propto\;
(\Delta x)^{2n_{\rm P}-n_{\rm L}(d-2)}
\;\longrightarrow\;0
\qquad \text{as } \Delta x\to0.
\end{equation}
Consequently, all non-Gaussian contributions vanish in the continuum
limit, leaving only the free propagator, possibly shifted by finite mass
or vacuum-energy counterterms.

This behavior is fully consistent with the well-known triviality of $\phi^4$ theory in
dimensions above four, where the Gaussian fixed point governs the continuum limit and no
interacting continuum fixed point exists.
While this analysis does not constitute a proof of triviality, it demonstrates that the
reorganized expansion introduced here respects the expected dimensional dependence and
reproduces the correct qualitative behavior of $\phi^4$ theory for $d>4$.
In contrast, for $d \le 4$ the above power counting no longer suppresses interaction
contributions in the continuum limit, indicating that the reorganized expansion retains
nontrivial dynamics in the physically relevant lower-dimensional cases.

\section{Conclusions and outlook}
\label{sec:conclusions}

In this work we have constructed a reorganized perturbative framework for scalar $\phi^{4}$
theory in which the quartic interaction is treated exactly and the kinetic operator is introduced
as a perturbation. Unlike the conventional weak-coupling expansion in powers of $\lambda$, which
is asymptotic and factorially divergent, the resulting kinetic-term expansion is term-by-term
convergent for all positive coupling at finite lattice spacing and finite volume.

The origin of this convergence is that the highest-order stabilizing interaction is never
expanded, but retained exactly in the exponential. As a result, the path integral is evaluated
against a measure with strong large-field suppression, and the perturbative expansion never
probes regions of field space that give rise to factorial growth. This mechanism is completely
transparent in the zero-dimensional model, where the reorganization leads to an exact series in
inverse fractional powers of $\lambda$ with coefficients controlled by Gamma functions. We proved
absolute convergence for all values of the mass and source and verified numerically that truncated
partial sums converge uniformly to the exact partition function and correlation functions.

We extended this construction to higher-dimensional $\phi^{4}$ theory by introducing auxiliary
variables that disentangle the kinetic term coupling $\phi_x$ and $\phi_{x'}$ at distinct
spacetime points. Once this entanglement is removed, the integration over the original field
becomes ultra-local and can be carried out exactly at each site. The resulting formulation yields
a systematic expansion in powers of the kinetic structure, with nontrivial interactions encoded
in explicitly known local vertices. On a discrete lattice, the expansion naturally organizes into
one-site, two-site, and higher multi-site contributions and is absolutely convergent at every
order. Explicit calculations on small lattices demonstrate rapid and systematic convergence
toward the exact lattice partition function as higher multi-site terms are included.

At the diagrammatic level, the auxiliary-field formulation leads to resummed propagators that
incorporate an infinite class of kinetic insertions. Higher-order loop diagrams are constructed
from these resummed propagators and local vertices. Power-counting analysis shows that in
dimensions $d>4$ interaction-induced corrections are suppressed in the continuum limit, in
accordance with the triviality of $\phi^{4}$ theory above four dimensions, while in lower
dimensions the expansion provides a controlled framework in which ultraviolet sensitivity is
isolated into a small number of resummed structures.

More broadly, the auxiliary-field formulation should be viewed as a general disentangling device,
rather than as a construction tied to any particular interaction order or to Gaussian auxiliary
measures. The essential requirement is the ability to reformulate the theory so that the original
fields can be integrated out exactly once the entanglement is removed; the auxiliary theory need
not be Gaussian, and the mechanism responsible for convergence does not rely on Gaussianity.

Several natural directions for future work follow from this perspective. A particularly
interesting application is two-dimensional $\phi^{4}$ theory, where the model flows to the
Ising conformal field theory at criticality. Extensions to theories with multiple interacting
fields, mixed kinetic structures, or vector-valued degrees of freedom would further clarify the
generality of the approach. On the lattice side, it would be valuable to explore larger systems
and to investigate whether truncated kinetic-term expansions can provide analytically controlled
approximations or improved starting points for numerical simulations.

Taken together, these results show that convergent perturbative expansions for interacting
quantum field theories can be obtained by reorganizing the path integral itself, provided the
highest-order stabilizing interactions are treated exactly. While this work focuses on scalar
$\phi^{4}$ theory, the underlying locality-first strategy appears to be considerably more
general and may offer a useful route toward controlled nonperturbative analyses of quantum
field theory.

\appendix

\appendix
\section{Details of the $0$-dimensional integrals}
\label{app:0d-details}

\subsection{Basic integral identities}

Throughout this work we repeatedly use the elementary identity
\begin{equation}
\int_{0}^{\infty} \dd\phi\; \phi^{p} e^{-\lambda \phi^{4}}
= \frac{1}{4}\,\lambda^{-\frac{p+1}{4}}
\Gamma\!\left(\frac{p+1}{4}\right),
\qquad p>-1,\ \Re\lambda>0,
\label{eq:basic-gamma}
\end{equation}
which follows immediately from the change of variables
$u=\lambda\phi^{4}$.

Integrals over the full real line reduce to \eqref{eq:basic-gamma}
by parity,
\begin{equation}
\int_{-\infty}^{\infty} \dd\phi\; \phi^{q} e^{-\lambda \phi^{4}}
=
\begin{cases}
0, & q \ \text{odd},\\[6pt]
2 \displaystyle\int_{0}^{\infty} \dd\phi\; \phi^{q} e^{-\lambda \phi^{4}},
& q \ \text{even}.
\end{cases}
\end{equation}

All local moments appearing in the $0$-dimensional expansion are
special cases of these integrals.

\subsection{Domain of validity}

All integrals in the $0$-dimensional theory are understood for
$\Re(\lambda)>0$, which guarantees absolute convergence of the defining
integrals due to the large-field suppression provided by the quartic
term.

Moments of the form $\int \dd\phi\, \phi^{p} e^{-\lambda\phi^{4}}$
are finite for $p>-1$ and are given explicitly by
Eq.~\eqref{eq:basic-gamma}.
Throughout the paper we only encounter nonnegative integer powers
$p$, so this condition is always satisfied.

Unless otherwise stated, statements of convergence and analyticity
are made at fixed $\lambda>0$, with $m^{2}$ and $J$ allowed to take
arbitrary complex values.

\subsection{Absolute convergence of the reorganized series}

Using the identities summarized above, the $0$-dimensional generating
functional may be written as a double series obtained by expanding the
Gaussian term,
\begin{equation}
Z(J)
=\frac{1}{2}\sum_{n,n'\ge 0}
\frac{(-1)^n}{n!\,(2n')!}
\left(\frac{m^{2}}{2}\right)^{n}
J^{2n'}\,
\lambda^{-\left(n+\frac{n'}{2}+\frac14\right)}
\Gamma\!\left(n+\frac{n'}{2}+\frac14\right),
\label{eq:0d-series-app}
\end{equation}
up to an overall $\lambda$-dependent normalization.

For fixed $\lambda>0$, the series \eqref{eq:0d-series-app} converges
absolutely for all complex $m^{2}$ and $J$.
Absolute convergence follows from a ratio test.
For fixed $n'$, define
\[
a_{n,n'}
\equiv
\frac{1}{n!\,(2n')!}
\left|\frac{m^{2}}{2}\right|^{n}
|J|^{2n'}
\lambda^{-\left(n+\frac{n'}{2}+\frac14\right)}
\Gamma\!\left(n+\frac{n'}{2}+\frac14\right).
\]
Then
\begin{equation}
\frac{a_{n+2,n'}}{a_{n,n'}}
=
\frac{1}{(n+1)(n+2)}
\left(\frac{|m^{2}|}{2\sqrt{\lambda}}\right)^{2}
\frac{\Gamma\!\left(n+\frac{n'}{2}+\frac54\right)}
{\Gamma\!\left(n+\frac{n'}{2}+\frac14\right)}
\;\xrightarrow[n\to\infty]{}\;0,
\end{equation}
using $\Gamma(x+1)=x\,\Gamma(x)$.
For fixed $n$, the sum over $n'$ is dominated by the factorial suppression
$(2n')!^{-1}$.

It follows that the series \eqref{eq:0d-series-app} converges absolutely
and uniformly on compact subsets of the $(m^{2},J)$ plane at fixed
$\lambda>0$.

\subsection{Uniform convergence and analyticity in $J$}

For fixed $\lambda>0$, the absolute convergence of the series
\eqref{eq:0d-series-app} implies uniform convergence on any compact
subset of the complex $J$ plane.
In particular, for bounded $|J|$ the sum over $n'$ is dominated by the
factorial suppression $(2n')!^{-1}$, while the sum over $n$ is controlled
by the ratio test discussed above.

As a consequence, the generating functional $Z(J)$ defines an entire
function of $J$ at fixed $\lambda>0$.
Derivatives with respect to $J$ may therefore be taken term by term,
and correlation functions obtained from $Z(J)$ are uniquely and
unambiguously defined by the reorganized expansion.

\subsection{Exchange of summation and integration}

Since the reorganized series \eqref{eq:0d-series-app} converges
absolutely for fixed $\lambda>0$, Fubini’s theorem applies and allows
the order of summation and integration in the $0$-dimensional path
integral to be exchanged.
All manipulations leading to the reorganized expansion, as well as the
extraction of correlation functions by differentiation with respect to
$J$, are therefore mathematically justified.

\section{Combinatorics of multi-site Gaussian moments}
\label{app:multisite}

In this appendix we record the general Wick-combinatorics formula used to evaluate
multi-site Gaussian moments of the auxiliary field.  Let $X$ be a centered Gaussian
vector with covariance $A$, i.e.
\begin{equation}
\frac{1}{\mathcal N}\int \prod_\ell \dd X_\ell\;
\exp\!\left[-\frac12\, X\cdot A^{-1}\cdot X\right]\,
X_{n}X_{m}
= A_{nm},
\qquad
\mathcal N \equiv \int \prod_\ell \dd X_\ell\;
\exp\!\left[-\frac12\, X\cdot A^{-1}\cdot X\right].
\end{equation}

For integers $a_i\ge 0$, define the normalized $k$-site moment
\begin{equation}
I^{(a_1,\dots,a_k)}_{n_1\cdots n_k}
\equiv
\frac{1}{(2a_1)!\cdots(2a_k)!\,\mathcal N}
\int \prod_\ell \dd X_\ell\;
e^{-\frac12 X\cdot A^{-1}\cdot X}\;
\prod_{i=1}^k X_{n_i}^{2a_i}.
\label{eq:Ik-def}
\end{equation}

\subsection{General closed form}

A convenient generating-function identity is
\begin{equation}
\frac{1}{\mathcal N}\int \prod_\ell \dd X_\ell\;
e^{-\frac12 X\cdot A^{-1}\cdot X + t\cdot X}
=
\exp\!\left(\frac12\, t^{\mathsf T} A\, t\right).
\end{equation}
Expanding the right-hand side and extracting the coefficient of
$\prod_{i=1}^k t_i^{2a_i}$ yields a compact general expression for
\eqref{eq:Ik-def}.  Introduce nonnegative integers $b_{ij}$ ($1\le i\le j\le k$),
where $b_{ii}$ counts self-pairings at site $n_i$ and $b_{ij}$ ($i<j$) counts
cross-pairings between $n_i$ and $n_j$.  These integers are constrained by
\begin{equation}
2b_{ii}+\sum_{j\ne i} b_{ij} = 2a_i,
\qquad i=1,\dots,k.
\label{eq:b-constraints}
\end{equation}
Let $\mathcal D(a_1,\dots,a_k)$ denote the finite set of all
$\{b_{ij}\}_{1\le i\le j\le k}$ satisfying \eqref{eq:b-constraints}.  Then
\begin{equation}
I^{(a_1,\dots,a_k)}_{n_1\cdots n_k}
=
\sum_{\{b_{ij}\}\in \mathcal D(a_1,\dots,a_k)}
\left[
\prod_{i=1}^k \frac{A_{n_i n_i}^{\,b_{ii}}}{2^{\,b_{ii}}\,b_{ii}!}
\right]
\left[
\prod_{1\le i<j\le k}
\frac{A_{n_i n_j}^{\,b_{ij}}}{b_{ij}!}
\right].
\label{eq:Ik-general}
\end{equation}

\subsection{Checks and special cases}

For $k=2$, the constraints imply $b_{12}=2b$ is even, and \eqref{eq:Ik-general}
reduces to
\begin{equation}
I^{(a_1,a_2)}_{nm}
=
\sum_{b=0}^{\min(a_1,a_2)}
\frac{A_{nn}^{\,a_1-b}A_{mm}^{\,a_2-b}A_{nm}^{\,2b}}
{2^{a_1+a_2-2b}\,(a_1-b)!\,(a_2-b)!\,(2b)!},
\end{equation}
in agreement with Eq.~\ref{eq:Inm-combinatorial}.  For $k=3$, \eqref{eq:Ik-general} reproduces
Eq.~\ref{eq:I123-combinatorial}.

\section{Numerical implementation details}
\label{app:numerical}

This appendix summarizes the numerical methods used to evaluate the truncated
strong-coupling expansions and correlation functions discussed in the main text.
The emphasis is on convergence, stability, and error control rather than on
language-specific implementation details.

\subsection{Evaluation of finite-dimensional integrals}

In the zero-dimensional model, all quantities of interest reduce to
one-dimensional integrals of the form
\begin{equation}
    Z(J) = \int_{-\infty}^{\infty} \dd \phi \,
    \exp\!\left( - \lambda \phi^4 - \frac{m^2}{2}\phi^2 + J\phi \right),
\end{equation}
as well as moments obtained by differentiation with respect to $J$.
These integrals were evaluated numerically using high-precision adaptive
quadrature over a finite domain, with the integration range chosen large
enough that the integrand was exponentially suppressed at the boundaries.
The quadrature tolerance was set such that further tightening had no
discernible effect on the quoted results.

Exact numerical quadrature results were used as benchmarks against which
truncated series expansions were compared.

\subsection{Truncated strong-coupling expansions}

The reorganized expansion derived in Sections~\ref{app:0d-details} and
\ref{sec:continuum-aux} involves sums over multi-site contributions and
internal combinatorial indices.
In practice, these sums were truncated at finite orders
\begin{equation}
    a \le a_{\max}, \qquad
    a' \le a'_{\max}, \qquad \ldots
\end{equation}
depending on the diagrammatic class under consideration.
Truncation orders were increased until successive partial sums differed
by less than a prescribed tolerance, typically at the level of
$10^{-6}$ or better.

For the zero-dimensional theory, convergence of the truncated series was
verified by direct comparison with the exact quadrature results.
In higher-dimensional examples, convergence was assessed by monitoring
the stability of correlation functions under increases of all independent
truncation parameters.

\subsection{Momentum-space integrals and lattice sums}

In the continuum theory formulated on a finite lattice,
momentum-space integrals were replaced by discrete sums over lattice momenta,
with ultraviolet convergence ensured by the finite lattice spacing.
All sums were evaluated using symmetric momentum grids, and convergence
was checked by increasing the lattice resolution and verifying that the
results remained unchanged within numerical precision.

Resummed auxiliary-field propagators, such as those discussed in
Appendix~\ref{app:effective-propagator}, were implemented analytically
as closed-form expressions obtained from geometric series, thereby
avoiding numerical instabilities associated with explicit summation
of self-energy insertions.

\subsection{Error estimates}

The dominant source of numerical uncertainty in all results is truncation
of the reorganized perturbative series rather than numerical integration
error.
This uncertainty was estimated by varying all truncation parameters
independently and taking the resulting spread as an error estimate.
Floating-point roundoff errors were found to be negligible at the working
precision used throughout.

All figures presented in the main text were generated only after
verifying numerical stability with respect to integration tolerances,
truncation orders, and lattice resolution.

\section{Resummation and the effective auxiliary-field propagator}
\label{app:effective-propagator}

As emphasized in Sec.~\ref{subsec:aux-expansion}, the auxiliary-field expansion is most naturally interpreted
in terms of resummed subclasses of diagrams rather than term-by-term evaluation.
The simplest and most important example is the resummation of repeated insertions of
the lowest-order local two-$X$ vertex, which reconstructs a finite effective
auxiliary-field propagator.

\subsection{Lowest-order local $X^2$ vertex}

Expanding the normalized local functional $\hat I(iX+J)$ (or equivalently $bI$ in the
main text) generates local vertices with $n_X$ insertions of $X(x)$.
The leading nontrivial vertex in the $X$-sector is the local two-$X$ insertion,
\begin{equation}
V_{X^2}(x)\;=\; -\,\frac{1}{2}\,
\frac{\Gamma(\tfrac34)}{\Gamma(\tfrac14)}\,
\Delta x^{\frac{d}{2}}\,\lambda^{-\frac12}\;X(x)^2,
\label{eq:V2X}
\end{equation}
where the overall normalization follows from the explicit $0$-dimensional integral
coefficients (cf.~App.~A).
Diagrammatically, $V_{X^2}$ may be viewed as a momentum-independent self-energy insertion.

\subsection{Geometric-series resummation}

The Gaussian measure for $X$ has covariance
\begin{equation}
\langle X(k)\,X(-k)\rangle_0 \equiv D_X^{(0)}(k) = k^2+m^2,
\end{equation}
since the auxiliary-field quadratic form is $X A^{-1} X$ with $A=(-\partial^2+m^2)$.

Note that because the auxiliary field $X$ appears with quadratic form $X A^{-1} X$, its Gaussian two-point function is proportional to $A$ rather than $A^{-1}$; equivalently, the bare propagator of $X$ is the inverse
of that of the original field $\phi$.

Repeated insertion of the local $X^2(x)$ vertex \eqref{eq:V2X}  into the auxiliary two-point function generates
the series shown in Fig.~\ref{fig:effective-aux-propagator},
\begin{equation}
D_X(k)
=
D_X^{(0)}(k)
+ D_X^{(0)}(k)\,\Sigma_0\,D_X^{(0)}(k)
+ D_X^{(0)}(k)\,\Sigma_0\,D_X^{(0)}(k)\,\Sigma_0\,D_X^{(0)}(k)
+\cdots,
\label{eq:Dyson-series}
\end{equation}
with $\Sigma_0 \equiv -V_{2X}$.
Since $\Sigma_0$ is momentum-independent, this is a geometric series and sums to
\begin{equation}
D_X(k)
=
\frac{D_X^{(0)}(k)}{1+\Sigma_0\,D_X^{(0)}(k)}
=
\frac{k^2+m^2}{1+\frac{1}{2}\frac{\Gamma(\tfrac34)}{\Gamma(\tfrac14)}
\Delta x^{\frac{d}{2}}\lambda^{-\frac12}(k^2+m^2)}.
\label{eq:DX-resummed-raw}
\end{equation}
This closed-form expression corresponds to the diagrammatic resummation shown in Fig.~\ref{fig:effective-aux-propagator}. 
\begin{figure}[t!]
\centering
\begin{tikzpicture}[>=Latex, line cap=round, line join=round, thick, scale=1.0, every node/.style={transform shape}]
\tikzset{
dot/.style={circle, fill=black, inner sep=1.4pt},
lbl/.style={font=\small},
eq/.style={font=\large},
dbl/.style={double, double distance=1.1pt, line width=0.8pt}
}

\node[lbl,align=center] at (0,2.1)
{$V_{X^2}=-\frac12\frac{\Gamma(\tfrac34)}{\Gamma(\tfrac14)}\Delta x^{d/2}\tilde\lambda^{-1/2} X^2$
\quad $\Sigma_0 = \frac12\frac{\Gamma(\tfrac34)}{\Gamma(\tfrac14)}\Delta x^{d/2}\tilde\lambda^{-1/2}$
};

\coordinate (a0) at (-6.4,0);
\coordinate (b0) at (-4.0,0);
\draw (a0) -- (b0);
\node[lbl,below] at (a0) {$X(x)$};
\node[lbl,below] at (b0) {$X(y)$};
\node[lbl,above] at ($(a0)!0.5!(b0)$) {$D_X^{(0)}(k)$};

\node[eq] at (-3.4,0) {$+$};

\coordinate (a1) at (-2.8,0);
\coordinate (b1) at (0.4,0);
\coordinate (m1) at ($(a1)!0.5!(b1)$);
\draw (a1) -- (b1);
\node[dot] at (m1) {};
\node[lbl,below] at (a1) {$X(x)$};
\node[lbl,below] at (b1) {$X(y)$};
\node[lbl,above] at ($(a1)!0.5!(m1)$) {$D_X^{(0)}(k)$};
\node[lbl,above] at ($(m1)!0.5!(b1)$) {$D_X^{(0)}(k)$};
\node[lbl,above] at ($(m1)+(0,0.65)$) {$V_{X^2}$};

\node[eq] at (1.0,0) {$+$};

\coordinate (a2) at (1.8,0);
\coordinate (b2) at (6.0,0);
\coordinate (m21) at ($(a2)!0.33!(b2)$);
\coordinate (m22) at ($(a2)!0.66!(b2)$);
\draw (a2) -- (b2);
\node[dot] at (m21) {};
\node[dot] at (m22) {};
\node[lbl,below] at (a2) {$X(x)$};
\node[lbl,below] at (b2) {$X(y)$};
\node[lbl,above] at ($(a2)!0.5!(m21)$) {$D_X^{(0)}(k)$};
\node[lbl,above] at ($(m21)!0.5!(m22)$) {$D_X^{(0)}(k)$};
\node[lbl,above] at ($(m22)!0.5!(b2)$) {$D_X^{(0)}(k)$};
\node[lbl,above] at ($(m21)+(0,0.65)$) {$V_{X^2}$};
\node[lbl,above] at ($(m22)+(0,0.65)$) {$V_{X^2}$};

\node[eq] at (6.6,0) {$+\cdots$};

\node[eq] at (-2.5,-1.15) {$=$};

\coordinate (ae) at (-1.9,-1.15);
\coordinate (be) at (1.9,-1.15);
\draw[dbl] (ae) -- (be);
\node[lbl,below] at (ae) {$X(x)$};
\node[lbl,below] at (be) {$X(y)$};
\node[lbl,above] at ($(ae)!0.5!(be)$) {$D_X(k)$};

\node[lbl,align=center] at (0,-2.8)
{$D_X(k)=\frac{D_X^{(0)}(k)}{1+\Sigma_0 D_X^{(0)}(k)}
=\frac{k^2+m^2}{1+\frac12\frac{\Gamma(3/4)}{\Gamma(1/4)}{\Delta x}^{d/2}\tilde\lambda^{-1/2}(k^2+m^2)}$};

\end{tikzpicture}
\caption{
Geometric-series resummation of repeated local $X^2$ insertions into the auxiliary-field two-point function
The local insertion vertex is $V_{X^2}=-\frac12\frac{\Gamma(\tfrac34)}{\Gamma(\tfrac14)}{\Delta x}^{d/2}\tilde\lambda^{-1/2}X^2$
}
\label{fig:effective-aux-propagator}
\end{figure}

It is convenient to package the coupling-dependent coefficient into the dimensionless
effective coupling introduced in Eq.~(56),
\begin{equation}
\tilde\lambda
\equiv
\lambda\,\Delta x^{-d}\left[\frac{2\Gamma(\tfrac14)}{\Gamma(\tfrac34)}\right]^2,
\qquad
\tilde\lambda^{-1/2}
=
\frac{1}{2}\frac{\Gamma(\tfrac34)}{\Gamma(\tfrac14)}\Delta x^{\frac{d}{2}}\lambda^{-1/2}.
\label{eq:lambdatilde-def-app}
\end{equation}
In terms of $\tilde\lambda$, \eqref{eq:DX-resummed-raw} becomes the compact form quoted
in the main text,
\begin{equation}
D_X(k)
=
\tilde\lambda^{1/2}
\left[
1-\frac{\tilde\lambda^{1/2}}{\tilde\lambda^{1/2}+m^2+k^2}
\right]
=
\tilde\lambda^{1/2}\,\frac{k^2+m^2}{k^2+m^2+\tilde\lambda^{1/2}}.
\label{eq:DX-resummed}
\end{equation}
For later use we also define the dimensionless factor
\begin{equation}
G(k)\equiv
1-\frac{\tilde\lambda^{1/2}}{\tilde\lambda^{1/2}+m^2+k^2}
=
\frac{k^2+m^2}{k^2+m^2+\tilde\lambda^{1/2}}
\label{eq:Gk-def-app}
\end{equation}
so that $D_X(k)=\tilde\lambda^{1/2}G(k)$.
Although Eq.~\eqref{eq:Dyson-series} is written as a geometric series, its radius
of convergence is limited by the condition $|\Sigma_0 D_X^{(0)}(k)|<1$.
Outside this domain, the term-by-term series representation ceases to converge.
Nevertheless, the auxiliary-field propagator is defined by its closed-form
expression Eq.~\eqref{eq:DX-resummed}, which is an analytic function of
$k^2+m^2$ and $\tilde\lambda^{1/2}$ for all $\lambda>0$.
In practice, the resummed propagator is understood as the analytic continuation
of the geometric series from the region where the latter converges.

\subsection{Exponential resummation of local vertex insertions}

The infinite site expansion Eq.~\ref{eq: site expansion} may be formally rewritten as an exponential,
\begin{equation}
\begin{aligned}
&1 +\int d^d x \, \mathcal{V}_x
+\frac{1}{2!}\int d^d x\, d^d y \, \mathcal{V}_x \mathcal{V}_y \\
&= \exp\Big[\int d^d x \, \mathcal{V}_x\Big]
\end{aligned}
\end{equation}
Combining this representation with the Gaussian term of the auxiliary field, the generating functional may be written in the compact form
\begin{equation}
\begin{aligned}
Z[J]
=
\int \mathcal{D}X \;
\exp\!\Big[
-\int d^dx \, X(\partial^2+m^2)^{-1}X
+\int d^d x \, \mathcal{V}_x[X,J]
\Big]
\label{eq:final form}
\end{aligned}
\end{equation}

The explicit form of the site-level vertex $\mathcal{V}_x$ is
\begin{equation}
\begin{aligned}
\mathcal{V}_x
&=
\sum_{n=1}^\infty
\frac{1}{(2n)!}
\frac{\Gamma(\frac{n}{2}+\frac{1}{4})}{\Gamma(\frac14)}
{\Delta x}^{\frac{(3n-2)d}{2}}\lambda^{-\frac{n}{2}}
\Big[ iX(x)+J(x) \Big]^{2n} \\
&=
\sum_{n=1}^\infty
\Bigg\{
\frac{(-1)^n}{(2n)!}
\frac{\Gamma(\frac{n}{2}+\frac{1}{4})}{\Gamma(\frac14)}
{\Delta x}^{\frac{(3n-2)d}{2}}\lambda^{-\frac{n}{2}}
X^{2n}(x)
\\
&-
\frac{i(-1)^n}{(2n-1)!}
\frac{\Gamma(\frac{n}{2}+\frac{1}{4})}{\Gamma(\frac14)}
{\Delta x}^{\frac{(3n-2)d}{2}}\lambda^{-\frac{n}{2}}
X^{2n-1}(x)J(x)
+\mathcal{O}(J(x)^2)
\Bigg\}
\end{aligned}
\end{equation}
Here the factor ${\Delta x}^{-d}$ implicit in $\mathcal{V}_x$ accounts for the conversion of the lattice sum over sites into a continuum integral.

The quadratic term in $X$ contained in $\mathcal{V}_x$ provides an alternative way to see the emergence of the effective auxiliary-field propagator, without reference to the radius of convergence of the original expansion. Explicitly,
\begin{equation}
\begin{aligned}
&-\int d^dx \Big\{
X(\partial^2+m^2)^{-1}X
-
\frac{1}{2!}
\frac{\Gamma(\frac34)}{\Gamma(\frac14)}
\tilde \lambda^{-\frac12}
X^2
\Big\}
\\
&=
-\int d^dk \;
X(k)
\Bigg[
\frac{
k^2+m^2
+
\frac{2\Gamma(\frac14)}{\Gamma(\frac34)}\tilde \lambda^{1/2}
}{
\big[2\frac{\Gamma(\frac14)}{\Gamma(\frac34)}\tilde \lambda^{1/2}\big]
[k^2+m^2]
}
\Bigg]
X(k)^*
\end{aligned}
\end{equation}

A remaining concern is whether the functional integral Eq.~\eqref{eq:final form} remains well defined in the joint limit of vanishing lattice spacing and infinite spatial volume. Formally, Eq.~\eqref{eq:final form} resembles a rewriting of the original $\phi^4$ theory, and it is not a priori obvious whether expanding the exponentiated vertex $\int d^dx\,\mathcal{V}_x$ and subsequently integrating over the auxiliary field $X$ yields a convergent series.

The essential distinction from the conventional weak-coupling expansion lies in the nature of the object whose powers are being summed. In the standard $\phi^4$ expansion one encounters moments of the form $\langle (\phi^4)^n\rangle_0$, whose Wick contractions grow factorially as $(2n)!$ and render the perturbative series asymptotic. In the present formulation, by contrast, the expansion is organized in powers of the site-level vertex $\mathcal{V}_x[X]$, whose coefficients are generated by a bounded Hubbard--Stratonovich profile function.

\begin{figure}[t]
\centering
\includegraphics[width=0.65\textwidth]{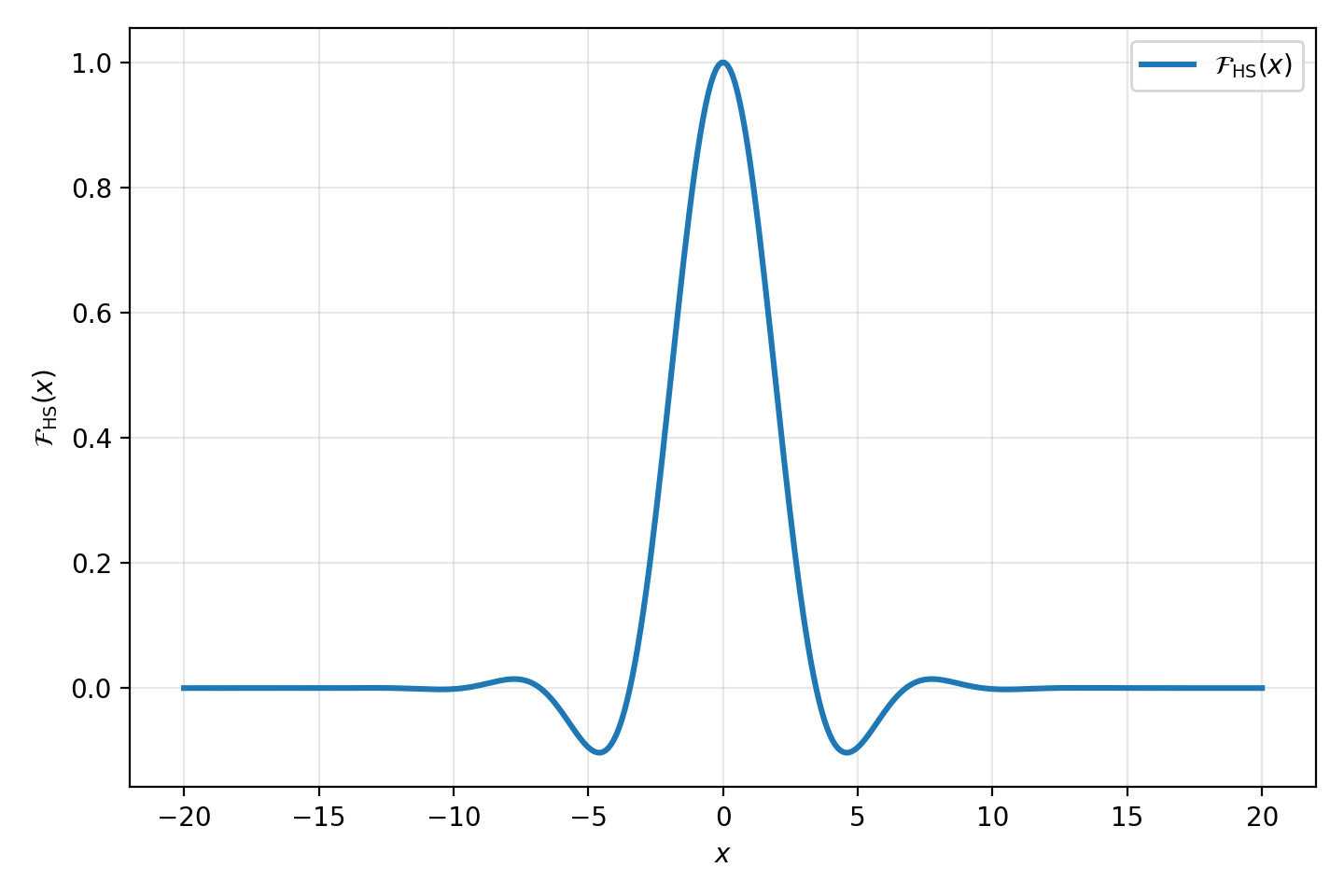}
\caption{
The Hubbard--Stratonovich profile function $\mathcal{F}_{\mathrm{HS}}(x)$
defined in Eq.~\ref{eq:f HS}
The bounded behavior for all real $x$ illustrates that the site-level
vertex entering the exponentiated representation does not exhibit
uncontrolled growth at large auxiliary-field amplitude
}
\label{fig:HS-profile}
\end{figure}

To isolate this point, it is useful to consider the dimensionless Hubbard--Stratonovich profile function
\begin{equation}
\mathcal{F}_{\mathrm{HS}}(x)
\equiv
\sum_{n=0}^\infty
\frac{(-1)^n}{(2n)!}
\frac{\Gamma(\frac{n}{2}+\frac14)}{\Gamma(\frac14)}
x^{2n}
\label{eq:f HS}
\end{equation}
This function remains bounded for all real $x$, as illustrated in Fig.~\ref{fig:HS-profile}. As a result, repeated insertions of the local vertex do not generate the factorial growth characteristic of moments of $\phi^4$.

At a heuristic level, this observation suggests that the expansion
\[
\sum_{n=0}^\infty \frac{1}{n!}\Big\langle \Big(\int d^dx\,\mathcal{V}_x[X]\Big)^n \Big\rangle_X
\]
is controlled by the boundedness of the site-level building block, rather than by cancellations among rapidly growing terms. While this argument does not constitute a proof of convergence of the full functional integral in the infinite-volume continuum limit, it explains why the present reorganization avoids the factorial growth responsible for the asymptotic nature of the conventional weak-coupling expansion.

\bibliographystyle{unsrt}
\bibliography{refs}

@article{dyson1952divergence,
  title={Divergence of perturbation theory in quantum electrodynamics},
  author={Dyson, Freeman J},
  journal={Physical Review},
  volume={85},
  number={4},
  pages={631},
  year={1952},
  publisher={APS}
}

@article{bender1969anharmonic,
  title={Anharmonic oscillator},
  author={Bender, Carl M and Wu, Tai Tsun},
  journal={Physical Review},
  volume={184},
  number={5},
  pages={1231},
  year={1969},
  publisher={APS}
}

@article{lipatov1977divergence,
  title={Divergence of the perturbation-theory series and pseudoparticles},
  author={Lipatov, LN},
  journal={JETP Lett.(USSR)(Engl. Transl.);(United States)},
  volume={25},
  number={2},
  year={1977},
  publisher={BP Konstantinov Institute of Nuclear Physics, USSR Academy of Sciences}
}

@article{seznec1979summation,
  title={Summation of divergent series by order dependent mappings: Application to the anharmonic oscillator and critical exponents in field theory},
  author={Seznec, R and Zinn-Justin, J},
  journal={Journal of Mathematical Physics},
  volume={20},
  number={7},
  pages={1398--1408},
  year={1979},
  publisher={American Institute of Physics}
}

@article{stevenson1981optimized,
  title={Optimized perturbation theory},
  author={Stevenson, Paul M},
  journal={Physical Review D},
  volume={23},
  number={12},
  pages={2916},
  year={1981},
  publisher={APS}
}

@book{kleinert2006path,
  title={Path integrals in quantum mechanics, statistics, polymer physics, and financial markets},
  author={Kleinert, Hagen},
  year={2006},
  publisher={World Scientific Publishing Company}
}

@article{BenderCooperGuralnikSharp1979,
  author    = {Carl M. Bender and Fred Cooper and Gerald S. Guralnik and David H. Sharp},
  title     = {Strong-coupling expansion in quantum field theory},
  journal   = {Phys. Rev. D},
  volume    = {19},
  pages     = {1865},
  year      = {1979}
}

@article{JankeKleinert1995,
  author    = {W. Janke and H. Kleinert},
  title     = {Variational perturbation expansion for strong-coupling coefficients of the anharmonic oscillator},
  journal   = {Phys. Rev. Lett.},
  volume    = {75},
  pages     = {2787},
  year      = {1995}
}

@article{Kleinert1998,
  author  = {Hagen Kleinert},
  title   = {Strong-coupling behavior of {$\phi^4$} theories and critical exponents},
  journal = {Phys. Rev. D},
  volume  = {57},
  pages   = {2264},
  year    = {1998}
}

@article{HamprechtKleinert2003,
  author  = {B. Hamprecht and H. Kleinert},
  title   = {Dependence of variational perturbation expansions on strong-coupling behavior: Inapplicability of {$\delta$}-expansion to field theory},
  journal = {Phys. Rev. D},
  volume  = {68},
  pages   = {065001},
  year    = {2003}
}

@article{SeroneSpadaVilladoro2018,
  author    = {Marco Serone and Gabriele Spada and Giovanni Villadoro},
  title     = {$\lambda \phi^4$ Theory I: The Symmetric Phase Beyond NNNNNNNNLO},
  journal   = {arXiv:1805.05882},
  year      = {2018}
}

@misc{MarinoResurgence,
  author    = {Marcos Mariño},
  title     = {An introduction to resurgence in quantum theory},
  note      = {lecture notes / review},
  year      = {2021}
}

@book{domb2000phase,
  title={Phase transitions and critical phenomena},
  author={Domb, Cyril},
  volume={19},
  year={2000},
  publisher={Elsevier}
}

@article{kogut1979introduction,
  title={An introduction to lattice gauge theory and spin systems},
  author={Kogut, John B},
  journal={Reviews of Modern Physics},
  volume={51},
  number={4},
  pages={659},
  year={1979},
  publisher={APS}
}

@article{benzi1978high,
  title={High temperature expansion without lattice},
  author={Benzi, R and Martinelli, G and Parisi, G},
  journal={Nuclear Physics B},
  volume={135},
  number={3},
  pages={429--444},
  year={1978},
  publisher={Elsevier}
}

@techreport{parisi1975high,
  title={High temperature expansion and the reggeon calculus},
  author={Parisi, G and others},
  year={1975},
  institution={Comitato Nazionale per l'Energia Nucleare, Frascati (Italy). Laboratori~…}
}

@article{svaiter2005strong,
  title={The strong-coupling expansion and the ultra-local approximation in field theory},
  author={Svaiter, NF},
  journal={Physica A: Statistical Mechanics and its Applications},
  volume={345},
  number={3-4},
  pages={517--537},
  year={2005},
  publisher={Elsevier}
}

@article{ignatyuk2022strong,
  title={Strong-weak Coupling Lattice Duality in Non-Local QFT with Application to Phase Transitions},
  author={Ignatyuk, Nikita A and Skliannyi, Daniel V},
  journal={arXiv preprint arXiv:2207.11503},
  year={2022}
}

@article{bernard2019nonlocal,
  title={Nonlocal scalar quantum field theory—Functional integration, basis functions representation and strong coupling expansion},
  author={Bernard, Matthew and Guskov, Vladislav A and Ivanov, Mikhail G and Kalugin, Alexey E and Ogarkov, Stanislav L},
  journal={Particles},
  volume={2},
  number={3},
  pages={385--410},
  year={2019},
  publisher={MDPI}
}

@article{kleinert2010converting,
  title={Converting divergent weak-coupling into exponentially fast convergent strong-coupling expansions},
  author={Kleinert, Hagen},
  journal={arXiv preprint arXiv:1006.2910},
  year={2010}
}

@article{li2011effective,
  title={Effective action approach to the p-band Mott insulator and superfluid transition},
  author={Li, Xiaopeng and Zhao, Erhai and Liu, W Vincent},
  journal={Physical Review A—Atomic, Molecular, and Optical Physics},
  volume={83},
  number={6},
  pages={063626},
  year={2011},
  publisher={APS}
}

@article{freericks1996strong,
  title={Strong-coupling expansions for the pure and disordered Bose-Hubbard model},
  author={Freericks, JK and Monien, H},
  journal={Physical Review B},
  volume={53},
  number={5},
  pages={2691},
  year={1996},
  publisher={APS}
}

@article{campostrini1995strong,
  title={Strong-coupling expansion of chiral models},
  author={Campostrini, Massimo and Rossi, Paolo and Vicari, Ettore},
  journal={Physical Review D},
  volume={52},
  number={1},
  pages={358},
  year={1995},
  publisher={APS}
}

@article{luscher1988application,
  title={Application of the linked cluster expansion to the n-component $\varphi$4 theory},
  author={L{\"u}scher, M and Weisz, P},
  journal={Nuclear Physics B},
  volume={300},
  pages={325--359},
  year={1988},
  publisher={Elsevier}
}

@article{butera2001remark,
  title={A remark on the numerical validation of triviality for scalar field theories using high-temperature expansions},
  author={Butera, P and Comi, M},
  journal={arXiv preprint hep-th/0112225},
  year={2001}
}

@article{gelfand1996series,
  title={Series expansions for excited states of quantum lattice models},
  author={Gelfand, Martin P},
  journal={Solid state communications},
  volume={98},
  number={1},
  pages={11--14},
  year={1996},
  publisher={Elsevier}
}

@article{wortis1974linked,
  title={Linked cluster expansion},
  author={Wortis, Michael},
  journal={Phase transition and critical phenomena},
  volume={3},
  pages={113},
  year={1974},
  publisher={Academic Press}
}

@article{procacci1998remark,
  title={A remark on high temperature polymer expansion for lattice systems with infinite range pair interactions},
  author={Procacci, Aldo},
  journal={Letters in Mathematical Physics},
  volume={45},
  number={4},
  pages={303--322},
  year={1998},
  publisher={Springer}
}

@article{reisz1995advanced,
  title={Advanced linked cluster expansion Scalar fields at finite temperature},
  author={Reisz, Thomas},
  journal={Nuclear Physics B},
  volume={450},
  number={3},
  pages={569--602},
  year={1995},
  publisher={Elsevier}
}

@article{chiku1998optimized,
  title={Optimized perturbation theory at finite temperature},
  author={Chiku, Suenori and Hatsuda, T},
  journal={Physical Review D},
  volume={58},
  number={7},
  pages={076001},
  year={1998},
  publisher={APS}
}

@incollection{jakovac2016optimized,
  title={Optimized Perturbation Theory},
  author={Jakov{\'a}c, Antal and Patk{\'o}s, Andr{\'a}s},
  booktitle={Resummation and Renormalization in Effective Theories of Particle Physics},
  pages={69--95},
  year={2016},
  publisher={Springer}
}

@article{guida1995convergence,
  title={Convergence of scaled delta expansion: anharmonic oscillator},
  author={Guida, Riccardo and Konishi, Kenichi and Suzuki, Hiroshi},
  journal={Annals of Physics},
  volume={241},
  number={1},
  pages={152--184},
  year={1995},
  publisher={Elsevier}
}

@article{yamada2007delta,
  title={Delta expansion on the lattice and dilated scaling region},
  author={Yamada, Hirofumi},
  journal={Physical Review D—Particles, Fields, Gravitation, and Cosmology},
  volume={76},
  number={4},
  pages={045007},
  year={2007},
  publisher={APS}
}

@article{sissakian1994variational,
  title={Variational perturbation theory},
  author={Sissakian, Alexey and Solovtsov, Igor and Shevchenko, Oleg},
  journal={International Journal of Modern Physics A},
  volume={9},
  number={12},
  pages={1929--1999},
  year={1994},
  publisher={World Scientific}
}

@article{sissakian1994ss,
  title={{\ss}-function for the $\phi$4-model in variational perturbation theory},
  author={Sissakian, AN and Solovtsov, IL and Solovtsova, OP},
  journal={Physics Letters B},
  volume={321},
  number={4},
  pages={381--384},
  year={1994},
  publisher={Elsevier}
}

@article{weissbach2002high,
  title={High-order variational perturbation theory for the free energy},
  author={Weissbach, Florian and Pelster, Axel and Hamprecht, Bodo},
  journal={Physical Review E},
  volume={66},
  number={3},
  pages={036129},
  year={2002},
  publisher={APS}
}

@article{hamprecht2004variational,
  title={Variational perturbation theory for summing divergent tunnelling amplitudes},
  author={Hamprecht, B and Kleinert, H},
  journal={Journal of Physics A: Mathematical and General},
  volume={37},
  number={35},
  pages={8561},
  year={2004},
  publisher={IOP Publishing}
}

@article{solovtsova2013variational,
  title={Variational perturbation theory and nonperturbative calculations in QCD},
  author={Solovtsova, OP},
  journal={Physics of Atomic Nuclei},
  volume={76},
  number={10},
  pages={1295--1300},
  year={2013},
  publisher={Springer}
}

@article{costin2019resurgent,
  title={Resurgent extrapolation: rebuilding a function from asymptotic data. Painlev{\'e} I},
  author={Costin, Ovidiu and Dunne, Gerald V},
  journal={Journal of Physics A: Mathematical and Theoretical},
  volume={52},
  number={44},
  pages={445205},
  year={2019},
  publisher={IOP Publishing}
}

@article{malbouisson1999non,
  title={A Non-perturbative Solution of the Zero-Dimensional lambda phi\^{} 4 Field Theory},
  author={Malbouisson, APC and Portugal, R and Svaiter, NF},
  journal={arXiv preprint hep-th/9909175},
  year={1999}
}

@book{le2012large,
  title={Large-order behaviour of perturbation theory},
  author={Le Guillou, Jean-Claude and Zinn-Justin, Jean},
  volume={7},
  year={2012},
  publisher={Elsevier}
}

@article{rivasseau2010loop,
  title={Loop vertex expansion for $\Phi$2k theory in zero dimension},
  author={Rivasseau, Vincent and Wang, Zhituo},
  journal={Journal of mathematical physics},
  volume={51},
  number={9},
  year={2010},
  publisher={AIP Publishing}
}

@article{rivasseau2018loop,
  title={Loop vertex expansion for higher-order interactions},
  author={Rivasseau, Vincent},
  journal={Letters in Mathematical Physics},
  volume={108},
  number={5},
  pages={1147--1162},
  year={2018},
  publisher={Springer}
}

@book{baez2014introduction,
  title={Introduction to algebraic and constructive quantum field theory},
  author={Baez, John C and Segal, Irving E and Zhou, Zhengfang},
  year={2014},
  publisher={Princeton University Press}
}

@incollection{jaffe2000constructive,
  title={Constructive quantum field theory},
  author={Jaffe, Arthur},
  booktitle={Mathematical physics 2000},
  pages={111--127},
  year={2000},
  publisher={World Scientific}
}

@article{sberveglieri2019renormalization,
  title={Renormalization scheme dependence, RG flow, and Borel summability in $\phi$ 4 Theories in d< 4},
  author={Sberveglieri, Giacomo and Serone, Marco and Spada, Gabriele},
  journal={Physical Review D},
  volume={100},
  number={4},
  pages={045008},
  year={2019},
  publisher={APS}
}

@incollection{rivasseau2000constructive,
  title={Constructive renormalization theory},
  author={Rivasseau, Vincent},
  booktitle={Particles and fields. Proceedings of the 10. Jorge Andre Swieca summer school},
  pages={389--416},
  year={2000}
}

@article{hancox2017solutions,
  title={Solutions in constructive field theory},
  author={Hancox-Li, Leif},
  journal={Philosophy of Science},
  volume={84},
  number={2},
  pages={335--358},
  year={2017},
  publisher={Cambridge University Press}
}

@article{sazonov2025variational,
  title={Variational Loop Vertex Expansion},
  author={Sazonov, Vasily},
  journal={Journal of High Energy Physics},
  volume={2025},
  number={4},
  pages={1--20},
  year={2025},
  publisher={Springer}
}

@article{seong2024central,
  title={Central limit theorem for the focusing $\Phi^{4}$-measure in the infinite volume limit},
  author={Seong, Kihoon and Sosoe, Philippe},
  journal={arXiv preprint arXiv:2411.07840},
  year={2024}
}

\end{document}